# From Influencers to Lecturers: Understanding Public Attitudes Toward Digital vs. Traditional Jobs

Paul H. P. Hanel[1], Gabriel Lins de Holanda Coelho[2], & Jennifer Haase[3]

[1]Department of Psychology, University of Essex, Colchester, United Kingdom, p.hanel@essex.ac.uk

[2]School of Applied Psychology, University College Cork, Cork, Ireland, linshc@gmail.com

[3]Weizenbaum-Institute, Berlin & Department of Computer Science, Humboldt University Berlin, Germany, jennifer.haase@hu-berlin.de

*Correspondence.* Correspondence should be addressed to: Paul Hanel, University of Essex, UK, p.hanel@essex.ac.uk

*Conflict of Interest.* The authors have no conflict of interest to declare. None of the studies was pre-registered.

*Acknowledgments.* This work was partly supported by the German Federal Ministry of Education and Research (BMBF), grant number 16DII133 (Weizenbaum-Institute). The responsibility for the content of this publication remains with the authors.

## Abstract

The rapid expansion of high-speed internet has led to the emergence of new digital jobs, such as digital influencers, fitness models, and adult models who share content on subscription-based social media platforms. Across two experiments involving 1,002 participants, we combined theories from social psychology and information systems to investigate how digital jobs are perceived compared to matched established jobs, and predictors of attitudes toward those jobs (e.g., symbolic threat, contact, perceived usefulness). We found that individuals in digital professions were perceived as less favorably and less hard-working than those in matched established jobs. Digital jobs were also regarded as more threatening to societal values and less useful. The relation between job type and attitudes toward these jobs was partially mediated by contact with people working in these jobs, perceived usefulness, perception of hard work, and symbolic threat. These effects were consistent across both experiments, and various moderators: openness to new experiences, attitudes toward digitalization, political orientation, and age. Among the nine jobs examined, lecturers were perceived as most positive, while adult models were viewed as least positive. Overall, our findings demonstrate that integrating theories from social psychology and information systems can enhance our understanding of how attitudes are formed.

# 1. Introduction

Over the past decades, and especially since the COVID-19 pandemic, the landscape of work has undergone a radical transformation. The necessity of remote work and digital resilience during the lockdowns brought digital tools into the mainstream, and allowed millions to work from home (Kumar et al., 2023). Yet, alongside this shift, digital professions have emerged – such as digital influencers, e-Sport players, OnlyFans models, and cryptocurrency investors – that remain controversial. Unlike traditional jobs that merely moved online (e.g., a lecturer teaching via Zoom), these new professions exist entirely within digital ecosystems.

Despite these roles becoming more common, anecdotal evidence suggests a deep-seated disregard for them, often characterized by the perception that they are not "real jobs." For instance, TikTok influencers have been called "stupid" and "entirely untalented" (Ritschel, 2022), OnlyFans models and travel influencers have been accused of not doing a proper job (DeSantis, 2019; Grant & McCallum, 2021) and e-Sport players have been described as too lazy to engage in "real manual labor" (BBC Radio 4, 2019). This creates a tension: digital skills are now seen as essential for keeping up professionally (Vivaldini & de Sousa, 2024), yet the individuals pioneering these new digital frontiers seem to be often stigmatized rather than celebrated.

At first glance, this dislike of digital jobs seems surprising given that a very similar version of these jobs exists in relatively more established areas. For example, the way digital influencers advertise products is often comparable to traditional forms of product placement or TV advertisements (Babin et al., 2021). Further, OnlyFans models who post adult content are comparable to people whose explicit pictures end up on tabloids or whose videotapes can be bought in adult stores. Moreover, frequent cryptocurrency investments have been associated with gambling (Mills & Nower, 2019), just as more established stock market trading has also been associated with gambling (Mosenhauer et al., 2021).

Drawing on theories from social psychology and information systems, we aim to provide a broader perspective on job perceptions that incorporates individual affective and cognitive processes. Social psychology offers valuable insights into how individuals perceive and feel toward other people. It provides theoretical frameworks to understand how societal norms and values influence perceptions of different kinds of jobs. In addition, we add theoretical perspectives from information systems research to elaborate on the specifics of *digitized* work. Specifically, in the present research, we investigate across two experiments (a) whether attitudes toward digital jobs are indeed more negative than toward established jobs, and (b) what underlying mechanisms such as contact with people in these jobs, symbolic threat, and perceived usefulness of these jobs may explain why attitudes toward digital jobs are more negative. Additionally, we separately test predictors of attitudes toward digital jobs. Below, we first discuss theories from social psychology and information systems, before we outline our hypotheses.

### 1.1. Social Psychological Theories

Over the past decades, several relevant theories have been proposed that explain how attitudes toward other people are formed: The theory of symbolic racism, integrated threat theory, and contact theory. The theory of symbolic racism originates from explaining subtle forms of prejudice toward black people in the USA (Kinder & Sears, 1981; Sears, 1988). For example, prejudice toward black people can be justified by claiming that they hold fewer Protestant values, such as hard work, which is used to explain the disadvantaged position of black people (Sears & Henry, 2003). Notably, the underlying assumption that perceived dissimilarity in values can lead to prejudice can be applied to groups beyond black people (Wolf et al., 2019).

Relatedly, integrated threat theory combines symbolic and realistic threats with intergroup anxiety and negative stereotypes (Stephan et al., 1999; Stephan & Stephan, 2000). Symbolic threats, which involve prejudice based on perceived differences in values or standards, are generally stronger predictors of attitudes than realistic threats, which concern the economic and political power of the ingroup. Notably, symbolic threat tends to be a stronger predictor of attitudes and prejudice than realistic threat (Gonzalez et al., 2008; Stephan et al., 1999). The latter finding has further been supported by the value-conflict hypothesis, which postulated and found that perceived value similarities predict attitudes toward outgroups even after controlling for prejudice-related dimensions such as racism, social dominance, and system justification (Chambers et al., 2013). This further supports the assumption that negative attitudes are at least partly based on perceived value conflicts.

Moreover, contact theory postulates that people have more negative views toward people with whom they have little or no contact (Allport, 1958). Conversely, people have more positive attitudes toward people they have more contact with, even if the contact is only digital (Costa et al., 2024; Schumann et al., 2017; Stiff & Bowen, 2016). This is partly because novel groups, which are typically outgroups, are more often associated with unique attributes, which tend to be perceived more negatively (Alves et al., 2018).

Those theories are empirically well supported (e.g., Gonzalez et al., 2008; Pettigrew & Tropp, 2006; Vedder et al., 2016). However, they have barely been used to predict attitudes toward people in specific jobs. It is therefore unclear whether those theories can be applied to predicting attitudes toward new digital jobs that have not existed when those theories were published. Additionally, and perhaps more intriguingly, it is unclear whether those social psychological theories can be advanced by combining them with theories from a relevant discipline, information systems.

### 1.2. Information System Theories

The theories from social psychology described above were mostly designed to explain attitudes and prejudice toward people who have a different ethnicity, immigration status, sexual orientation, gender, political orientation, or religious beliefs. While we believe that they can, in general, also be used to understand attitudes toward other groups of people, we argue that they might be missing an important predictor when it comes to understanding attitudes toward people working as digital influencers or online content subscription service models. Specifically, information systems as a field offers a useful perspective for understanding the role of technology and how its implementation and use within work contexts influence job structures, roles, as well as the nature of work itself (Vazquez et al., 2019). In the context of digital jobs, it can provide a nuanced understanding of the opportunities and challenges posed by digitalization (Legner et al., 2017), the competencies required in digital jobs, and the ways in which these jobs interact with and reshape existing organizational structures and practices (Soto-Acosta, 2020; Wibowo et al., 2022).

In the context of a rapidly evolving technological landscape, digital work has emerged as a distinctive domain that is characterized by unique attributes distinguishing it from more traditional forms of work. The digital nature of such work primarily refers to the fact that the tasks, processes, and communications involved are mediated by digital technologies and platforms (Baptista et al., 2020) and are fast-paced (Huang et al., 2017). These technologies enable work to be performed irrespective of physical location, contributing to telework or remote work, allowing for flexible working hours, and providing the workers with a sense of autonomy. The digital aspect profoundly changes the social dynamics of work. While offering the possibility of global collaboration, it also challenges traditional concepts of team cohesion, workplace culture, and presents challenges for leadership in digital contexts (Kane et al., 2019; Larson & DeChurch, 2020).

One established theoretical lens to study the application and efficacy of technology comes with the technology acceptance model (Davis, 1989; Venkatesh et al., 2003). While there is a debate about the usefulness of the various variants of technology acceptance model (Benbasat & Barki, 2007; Chuttur, 2009), we are mostly interested in one variable of the model that is a sub-dimension of performance expectancy (Venkatesh et al., 2003): Usefulness.

Usefulness was originally defined as "the degree to which a person believes that using a particular system would enhance his or her job performance" (Davis, 1989, p. 320). Perceived high levels of usefulness are positively associated with attitudes toward information systems and IT innovations and their usage (Adams et al., 1992; Dwivedi et al., 2019). We abstract this concept by applying it to the usefulness of digital jobs and the people doing these jobs, in contrast to previous research that primarily applied usefulness to technology acceptance. In other words, we aim to understand whether the perceived value (i.e., usefulness) of a profession influences attitudes toward individuals working in that field, rather than focusing on the underlying technologies of the profession itself. Thus, even though usefulness in Information Systems refers to the value of a technological tool for completing a task, and our research reframes it as the value of a profession for society, both approaches converge on the idea that perceptions of usefulness shape attitudes by signalling whether something is seen as valuable and worth engaging with.

We believe that it is justified to apply usefulness to jobs and the people doing those jobs, as expressions such as "what you are doing for work is extremely important", "thank you for your service", "you are important" (or, on the flip side, "you are useless"), "you are a valuable member of this community/organisation" are common expressions to signal that a specific job or person is perceived to be useful. Those theoretical considerations are also supported by the recent social utility-based acceptance/rejection model (Dambrun, 2024), which predicts that attitudes toward people are, at least partly, shaped by their perceived social utility. People who contribute to society are perceived more positively than those who are seen as a burden (e.g., those receiving social benefits). This model has, to the best of our knowledge, not been empirically tested yet.

Furthermore, other related theories such as job characteristics theory (Hackman & Oldham, 1976, 1980) or relevance theory (Schulz, 2003) do not contain a variable that directly expresses usefulness. We return to them in the General Discussion. Moreover, while psychological theories such as the theory of planned behavior (Ajzen, 1991) initially contributed to shaping information systems theories (Dwivedi et al., 2019), we argue that theories from information systems can also enrich psychological theories.

**1.3. The Present Research**

In the present research, we postulate that people pursuing digital jobs are perceived as less favorable, less hard-working, more of a threat to people's values, and that their work is perceived as less useful compared to people who are pursuing a more established job. This is because digital jobs are more novel, meaning people will interact less with them and perceive them as more negative (Alves et al., 2018; Pettigrew & Tropp, 2006). Crucially, we propose distinct theoretical pathways for this negativity. While Symbolic Threat captures the value-based conflict – the feeling that digital workers are violating traditional Protestant work ethics – Perceived Usefulness captures an instrumental judgment. Even if a digital job is seen as "hard work" (low symbolic threat), it may still be rejected if it is perceived as contributing nothing of value to society (low usefulness). This distinction is vital in the post-pandemic context, where digital tools are valued for their utility (Kumar et al., 2023), but the social utility of the "influencer economy" remains contested. Furthermore, people can be skeptical of IT-related changes (Laumer & Eckhardt, 2010). The high flexibility and fleetingness associated with digital technologies and trends make their development less tangible and credible. Following the optimal distinctiveness model (Brewer, 1991), people might be motivated to perceive digital jobs more negatively because it helps them to maintain a positive social identity (Tajfel & Turner, 1979).

Additionally, we investigate whether usefulness can also predict attitudes toward people doing a specific job, thereby going beyond social psychological research that has not included usefulness to the best of our knowledge. This omission might not be surprising given that usefulness is, in general, something that comes to mind when thinking about technological innovations or systems rather than people doing specific jobs (Davis, 1989). However, when a person doing a specific job is assessed, it might be perceived in relation to the technology associated with that profession (Wolfe & Patel, 2019) or its perceived usefulness to society. This prediction is supported by research which found that perceiving immigrants as contributing economically and symbolically to society (i.e., if immigrants are perceived as more useful) is associated with more positive attitudes (Li & Kung, 2023). We therefore predict that the perceived usefulness of a digital job is significantly associated with people's attitudes toward those who occupy such roles. Usefulness, in our view, is conceptually independent from realistic and symbolic threats, because something useful can be a threat to our economic system or values, but can also be an asset. We also test whether usefulness is an independent predictor of attitudes after controlling for variables from social psychological theories, such as symbolic threat and contact. This approach could advance both social psychological and information systems theories by providing a deeper understanding of the factors that influence attitudes toward individuals in specific jobs

By integrating perceived usefulness from information systems, specifically the technology acceptance model (Venkatesh et al., 2003) with symbolic threat (Stephan & Stephan, 2000) and contact (Pettigrew & Tropp, 2006) theories from social psychology, we seek a better understanding of attitudes toward digital jobs, beyond what each field would predict alone. Theories from social psychology primarily account for attitudes driven by social and symbolic dynamics, while information systems theories emphasize functional value. Synthesizing these perspectives enables us to investigate how instrumental judgments (usefulness) and social judgments (symbolic threat, contact) jointly influence attitudes. This combined approach could reveal how both practical and social evaluations contribute to the perception of digital jobs, thereby advancing theories in both fields.

To test our main predictions, we selected nine digital jobs (i.e., professions that have only recently emerged and are dependent on a digital infrastructure) for which we could identify a comparable, more established job. For example, we described the job of an interviewer in our experiments as either a talk-show host who invites celebrities from different areas, including actors, politicians, artists, and scientists, for interviews (established job), or someone who is also interviewing the same group of people, but broadcasting the interviews themselves across various platforms such as Spotify and YouTube (digital job).

We also hypothesize that this association between types of jobs (established vs. digital) is mediated by contact and perceived usefulness (Experiment 1) as well as contact, perceived usefulness, perception of hard work, and symbolic threat (Experiment 2). We expect, as discussed, that digital jobs are differently rated on all those variables than established jobs (e.g., digital jobs will be perceived as less useful). As outlined above, all mediators were found, individually, to be associated with attitudes towards various groups of people (e.g., Chambers et al., 2013; Costa et al., 2024; Gonzalez et al., 2008; Sears & Henry, 2003). Hence, contact, perceived usefulness, perception of hard work, and symbolic threat should mediate the association between job type and attitudes. This design simultaneously allowed us to test predictors of attitudes toward digital jobs. Finally, we tested across both experiments whether openness to new experiences, attitudes toward digitalization, political orientation, and age moderated any of the effects of condition (digital vs. established jobs).

We pondered that the effect of conditions might be larger for less open-minded people, who had fewer positive attitudes toward digitalization, were more conservative, and were older. For example, people with more negative attitudes toward digitalization might perceive digital jobs disproportionately negatively, whereas this is less the case for people holding more positive views toward digitalization.

## 2. Experiment 1

### 2.1. Method

Our experiments were approved by the local ethics committee at the institution of one of the authors. The data, full surveys, R-code to reproduce our analyses, and supplemental materials can be found here https://osf.io/wh2a3/?view_only=b1c267af87184f29949b591e52a75ed9.

**2.1.1. Participants.** A power analysis revealed that to detect a small effect size of $f$ = 0.125 in a 2 x 9-mixed-design with a power of .95, 464 participants are required. Our sample consisted of 502 participants living in the UK ($M_{age}$ = 45.96, $SD_{age}$ = 15.81, 106 participants were between 18-29 years old, 115 were 30-44, 214 were 45-64, and 63 were 65 years or older; 246 women, 247 men, 3 others, 2 prefer not to say, 4 participants did not respond to this item). Half of our participants were 50 years or older to ensure a wider age-spread. The majority of participants reported that their household income is <£25,000 ($n$ = 145) or £25,000-50,000 ($n$ = 197). A total of 65 participants had no formal degree, 155 had a high school degree, 195 had a bachelor's degree, 68 had a master's degree, and 16 had a Ph.D. (3 missing responses). Eighty-six participants owned cryptocurrencies (e.g., Bitcoins), 409 did not, and 4 did not know what cryptocurrencies were (3 missing responses). We recruited participants online on Prolific Academic (prolific.com) and compensated them with the living wage (pro rata) as suggested by Prolific.

**2.1.2. Design.** We used a 2 (between-subject factor: established vs. digital jobs) × 9 (within-subject: 9 jobs) mixed-design.

**2.1.3. Materials and Procedure.** After providing informed consent, participants were randomly allocated into either the digital or established jobs condition. We designed nine pairs of vignettes, each describing a similar job, with one job being digital and the other a similar established version of it (see Table S1 in the Supplemental Materials). The jobs were selected based on their prominence and whether it was possible to find a comparable established job in which the core functions would be as analogous as possible. We employed a specific set of a priori matching criteria to ensure the comparability of the established and digital job pairs. First, both jobs in a pair had to share a primary function (e.g., "interviewing" for podcaster/talk-show host; "investing" for crypto/stock market). Second, both had to offer comparable autonomy and potential for high income, ensuring that any differences in attitude could not be attributed to class or status differences alone. Third, the "established" version had to be a role that existed and was recognized well before the advent of social media, whereas the "digital" version had to be native to online platforms. We define digital jobs as work that is primarily conducted through online platforms, where individuals generate income by providing services, content, or expertise to a digital audience. The digital job would need to be mostly done remotely online and have emerged more widely after the established job. These roles typically involve capitalizing on digital media or technologies to connect with and deliver value to others. The established jobs would have been traditionally in-person but may have transitioned to online work more recently (e.g., particularly after the COVID-19 pandemic). We argue the digital and established version of each job can be meaningfully compared because there are parallels in job functions (e.g., professional competition in sports), in income (we added to each vignette that the job allowed them to have a good life), and professionalism (a good understanding of the tools used and market segment seems necessary for all jobs).

The nine selected jobs were (digital/established version) e-football player/footballer, Youtuber/TV presenter, digital influencer/advertising, OnlyFans model/adult model, workout tips on social media/personal trainer, podcaster/talk-show host, bitcoin investor/stock market investor, online psychotherapist/psychotherapist, and online-only lecturer/lecturer. For example, the established version of podcaster/talk-show read, "*Paul has a solid career as a talk-show host: He has a famous TV show on a large news channel, and interviews celebrities from different areas, including actors, politicians, artists, and scientists. Paul's income allows him to have a good life*." The digital version read "*Paul has a solid career as a podcaster: He has integrated channels on multiple platforms (e.g., Spotify, Deezer, Youtube), and interviews celebrities from different areas, including actors, politicians, artists, and scientists. He then shares the recorded interviews on the platforms. Paul's income allows him to have a good life*." We did not provide any salary estimates, because these substantially differ within- and between-countries, as well as between employers. We presented vignettes in randomized order.

Following each vignette, we assessed participants' attitudes toward the job and the person doing the job with seven items. Items were adapted from the literature (e.g., Armitage et al., 1999) but also based on derogative comments used by people when referring to digital jobs (e.g., "*X is not a real job*"). Example items include "*Do you consider what <name> is doing a real job*?", "*How favorable do you feel toward this job*?", and "*How much creativity does a <job title> require*?". All items were answered on 7-point response scales ranging from 1 (Definitively not/Not at all/None) to 7 (Definitively yes/Very/A great deal). A maximum likelihood factor analysis performed with the R-package nFactors (Raiche & Magis, 2022; version 2.4.1.1) resulted in a one factor solution (one eigenvalue > 1).

Additionally, the Kaiser-Gutmann Criterion, Optimal Coordinates, and visual inspection of the screeplot suggested one factor. We therefore averaged across all seven items (α = .89).

Usefulness was measured with one item: "*How much value does <name`s> work add to society*?" Responses were also given on a 7-point scale ranging from 1 (None) to 7 (A great deal).

We measured contact with the job (e.g., "*In the past year, how often have you listened to <job title>?*"). Responses were given on an 8-point scale ranging from 1 (Never) to 8 (A couple of times daily).

We measured the personality trait openness to new experience with the four-item version proposed by de Vries (2013). Example items include "*I have a vivid imagination*" and "*I am not interested in abstract ideas*" (reversed scored, α = .75).

Attitudes toward digitalization were measured with two identical items (adapted from Armitage et al., 1999) which read "*I see the internet and digitalization of our lives as*" but that had different response scales (1: Undesirable to 7: Desirable and 1: Bad to 7: Good, $r$ = .86).

Political orientation was measured with two items (adapted from Zarzeczna et al., 2024) "*Where on the following scale would you place your political ideology?*" that had different response scales (1: Very liberal to 7: Very conservative, 1: Very left wing to 7: Very right wing, $r$ = .88).

**2.2. Results**

To test whether attitudes toward some jobs were more positive, whether the job was described as established or digital, and whether there was an interaction between the two factors, we ran a 2x9-mixed ANOVA (Table S2 for descriptive statistics and effect sizes). As an effect size, we report the generalized eta-square $\hat{\eta}_G^2$ because it allows comparisons of a wider range of designs than (partial) eta-square (Olejnik & Algina, 2003). The main effect of condition, $F(1, 499) = 70.24$, $p < .001$, $\hat{\eta}^2_G = .07$, the main effect of job, $F(6.65, 3318.40) = 423.76$, $p < .001$, $\hat{\eta}^2_G = .29$, and the interaction, $F(6.65, 3318.40) = 20.85$, $p < .001$, $\hat{\eta}^2_G = .02$, were significant (Figure 1). Simple main effects of condition revealed that participants perceived all established (vs. digital) jobs more positively except for interviewer, for which the mean difference was not significant (Bonferroni corrected). Simple main effects of job revealed that participants perceived models as least favorable and lecturers as most favorable. To illustrate this pattern with a diagnostic example: Consider the pair of established vs digital advertiser. Despite both roles involving the creation of promotional content for products, participants rated the established advertiser significantly more favorably ($M = 5.51$, $SD = 0.85$) than the digital advertiser ($M = 4.42$, $SD = 1.34$), $p < .001$, and perceived the established advertiser's work as contributing significantly more value to society ($M = 4.12$, $SD = 1.49$, vs. $M = 3.24$, $SD = 1.69$, $p < .001$; see Table S2 for additional inferential statistics and effect sizes). This specific contrast highlights that the medium of the work (social media vs. traditional media) drives a "penalty" in attitudes and perceived usefulness.

The pattern for perceived usefulness of the job as dependent variable was similar (Figure S1): Participants found established jobs more useful, $F(1, 498) = 39.65$, $p < .001$, $\hat{\eta}^2_G = .03$, perceived the jobs differently in their usefulness, $F(7.24, 3605.18) = 584.64$, $p < .001$, $\hat{\eta}^2_G = .41$, which in turn also dependent on the condition (i.e., significant interaction), $F(7.24, 3605.18) = 26.27$, $p < .001$, $\hat{\eta}^2_G = .03$. The pattern for contact as dependent variable differed from perceived usefulness: While the main effect of condition, $F(1, 499) = 108.25$, $p < .001$, $\hat{\eta}^2_G = .05$, the main effect of job, $F(5.87, 2929.80) = 279.24$, $p < .001$, $\hat{\eta}^2_G = .30$, and the interaction, $F(5.87, 2929.80) = 73.73$, $p < .001$, $\hat{\eta}^2_G = .10$, were significant, contact was highest with presenter (media) and lowest with model and psychotherapists (Figure S2). On average, participants had more contact with established jobs.

**Figure 1**

*Attitudes toward a job depend on the type of job and condition (established vs. digital). Error bars represent 95%-Cis (Experiment 1).*

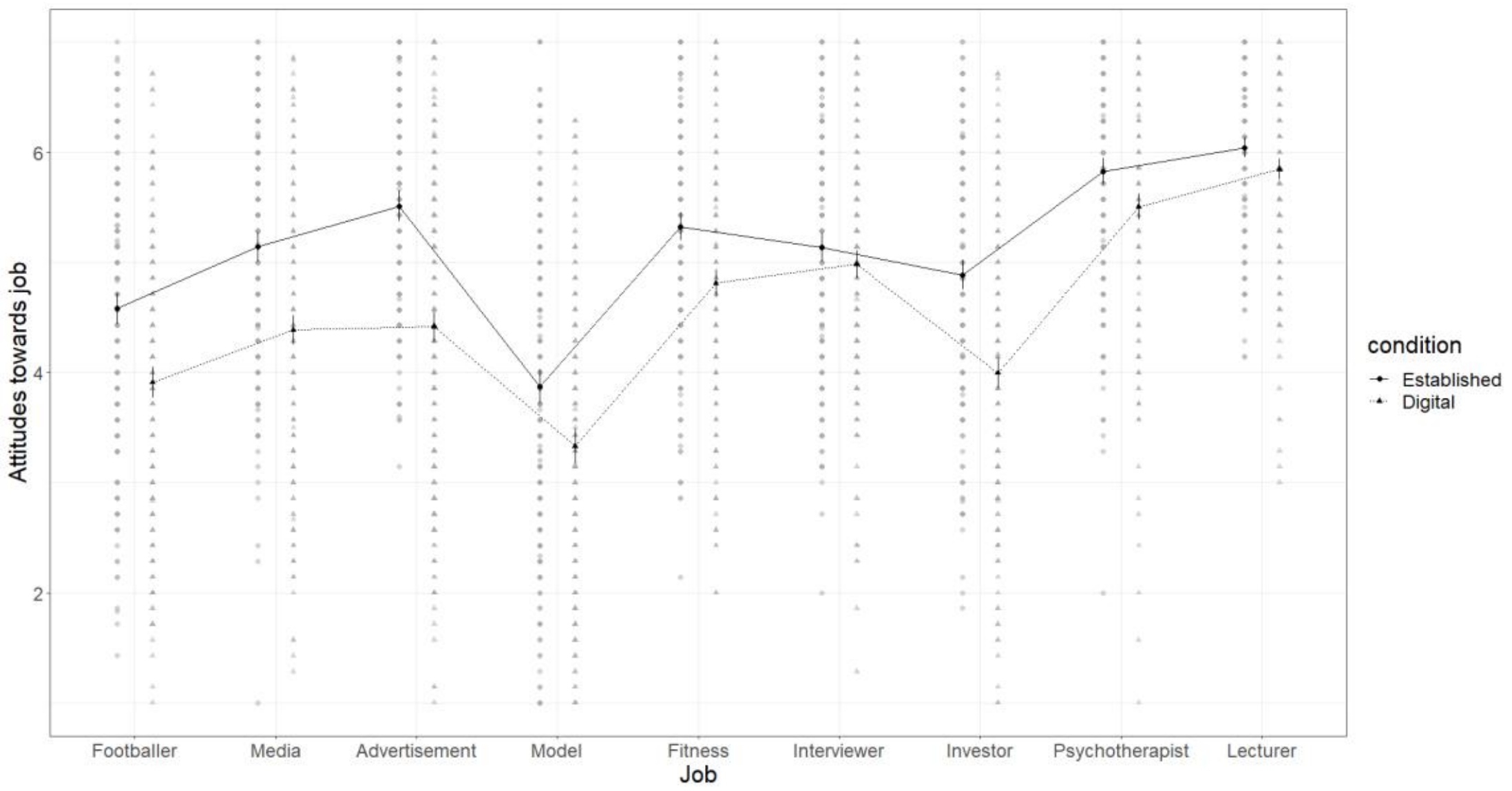

To test whether usefulness and contact mediate the association between condition and attitudes toward the job, we ran nine mediation models, one for each job (see Table S3 for total, direct, and indirect effects). The indirect effect was significant (i.e., the 95%-CI did not include zero) for all professions except fitness instructor, interviewer, and psychotherapist. Exploratory follow-up analyses revealed that both mediators – usefulness and contact – independently functioned as partial mediators.

To test whether openness to new experiences, attitudes toward digitalization, political orientation, or age moderated any of the effects of condition (digital vs. established jobs), we ran four linear mixed-effects models with condition and one moderator at a time as predictors, along with their interactions (Table S4 for correlation coefficients). We used linear mixed-effects models because they allow to generalize across stimuli (Judd et al., 2012). Attitudes toward the job was the outcome variable. Given that we ran four tests, we set the alpha-threshold to .05/4 = .0125 (i.e., Bonferroni correction). In other words, for this exploratory analysis we only interpret findings with $p < .0125$ as statistically significant.

When we included openness as a moderator, the two-way interaction was not significant, $B = -.05$, $SE = .06$, $p = .417$, neither was the effect of condition, $B = -.34$, $SE = .31$, $p = .277$. However, openness significantly predicted attitudes toward jobs, $B = .11$, $SE = .04$, $p = .007$. Exploratory simple slope analysis revealed that the association between openness and attitudes was only significant for presenter (media) and interviewer (Figure S3). This indicates that people who are more open, have more positive attitudes toward these two jobs.

When attitudes toward digitalization were included as a moderator, the two-way interaction was not significant at the adjusted alpha-level of .01, $B = .11$, $SE = .05$, $p = .029$, but the two main effects of condition, $B = -1.06$, $SE = .26$, $p < .001$ and attitudes toward digitalization, $B = .12$, $SE = .03$, $p < .001$, reached statistical significance. Exploratory simple slope analysis revealed that the association between attitudes toward digitalization and attitudes toward the job was significant and positive for all jobs except model and fitness instructor (Figure S4).

When political orientation was included as a moderator, neither the two-way interaction was significant, $B = .01$, $SE = .05$, $p = .816$, nor the main effect of political orientation, $B = -.07$, $SE = .04$, $p = .122$. Only the main effect of condition reached statistical significance, $B = -.60$, $SE = .21$, $p = .005$. Exploratory simple slope analysis revealed that the association between political orientation and attitudes toward the job was only significant for the job model across both conditions: More conservative people expressed less positive attitudes toward these two jobs (Figure S5).

When age was included as a moderator, neither the two-way interaction was significant, $B > -.01$, $SE < .01$, $p = .978$, nor the main effect of condition, $B = -.56$, $SE = .26$, $p = .065$. Only the main effect of age reached statistical significance, $B = -.01$, $SE < .01$, $p < .001$, with older people generally expressing more negative attitudes toward the jobs, independent of the condition. Exploratory simple slope analysis revealed that the association between age and attitudes toward the job was significantly negative for all jobs, but was only significant for advertisement, interviewer, and model across both conditions (Figure S6).

Finally, we tested whether any of the mediators, moderators, or demographic variables would uniquely predict attitudes toward jobs separately for each condition (established vs digital jobs), using linear-mixed effects models. Interestingly, usefulness and contact were the strongest predictors of attitudes toward both digital and established jobs, followed by attitudes toward digitalization and openness (Table 1). Also, as one might expect, political orientation was negatively associated with attitudes toward digital job, suggesting that left-wingers hold on average more positive views than right-wingers. Perhaps surprisingly, age did not predict attitude toward digital jobs.

**Table 1**

*Predictors of attitudes toward job separately for each condition (Experiment 1)*

| | Digital jobs | | | Established jobs | | |
|---|---|---|---|---|---|---|
| | *B* | *SE* | *p* | *B* | *SE* | *p* |
| Usefulness | 0.52 | .01 | <.001 | 0.44 | .01 | <.001 |
| Contact | 0.06 | .01 | <.001 | 0.07 | .01 | <.001 |
| Openness | 0.05 | .03 | .009 | 0.06 | .03 | .020 |
| Attitudes toward digitalization | 0.08 | .03 | .002 | 0.07 | .02 | .001 |
| Political orientation | -0.06 | .03 | .013 | -0.03 | .02 | .126 |
| Age | -0.03 | .00 | .254 | -0.01 | .00 | .007 |
| Gender (1: men, 2: women) | 0.03 | .07 | .672 | 0.03 | .06 | .658 |
| Education level | -0.05 | .04 | .166 | -0.04 | .03 | .236 |
| Income | -0.00 | .03 | .942 | 0.03 | .03 | .294 |

## 3. Experiment 2

Experiment 1 found that people tend to have more positive attitudes toward more established jobs than similar digital jobs. This association was partially mediated by usefulness and contact, which were also reliable predictors of attitudes toward digital jobs.

However, our attitude measure was broad, and our measures of contact and usefulness consisted of single items. While single items are shown to often be robust (Himmels et al., 2025; Sibley et al., 2024), in Experiment 2, we measured all constructs using multiple items. This also allowed us to test whether our findings replicate across different operationalizations of the same construct. Additionally, we included measures of symbolic threat and how hard-working people doing specific professions are perceived to test predictions from the integrated threat theory (Stephan & Stephan, 2000).

**3.1. Method**

**3.1.1. Participants.** Since we used the same 2 (condition) x 9 (jobs) mixed design as in Experiment 1, we aimed to collect a similar number of participants. In total, we recruited 500 participants living in the UK ($M_{age}$ = 47.15, $SD_{age}$ = 13.77, 55 participants were between 18-29 years old, 163 were 30-44, 220 were 45-64, and 56 were 65 years or older; 246 women, 247 men, 1 other, 6 participants did not respond to this item). The majority reported that their household income is <£25,000 ($n$ = 129) or £25,000-50,000 ($n$ = 211; in British Pounds). Fifty-one participants had no formal degree, 166 had a High-school degree, 203 had a Bachelor's degree, 65 had a Master's degree, and 11 had a PhD (4 missing responses to this item). Seventy-two participants owned cryptocurrencies (e.g., Bitcoins), 424 did not, and 2 did not know what cryptocurrencies were (2 missing responses). To further increase the age spread, half of the invited participants were at least 50 years old. This was done because age was found to be negatively associated with modern technology (Schönmann et al., 2024).

**3.1.2. Design.** We again used a 2 (between-subject factor: established vs. digital jobs) × 9 (within-subject: 9 jobs) mixed-design.

**3.1.3. Materials and Procedure.** The procedure was the same as in Experiment 1. We also used the same vignettes, except that we changed the gender of the person described in the vignettes to test whether the findings are independent of the person's gender described in the vignettes (i.e., not based on gender stereotypes). That is, the gender pronouns and names were changed from male to female or vice versa. The dependent variable and the mediators were measured with two items to avoid making the survey overly long.

Attitudes were measured with one item derived from the theory of planned behavior (Ajzen, 1991) “*How favorable do you feel toward this job*?” and one derived from the competence dimension of the stereotype content model (Fiske et al., 2002) “*How competent is <name>?”.* The two items were highly correlated, $r = .59$. However, even though both items are strongly correlated, they tap into slightly different aspects. We therefore re-ran our analyses with each item separately but we were able to replicate our findings, indicating the robustness of our findings.

Perceived usefulness was measured “*How useful is what <name> does*?” and “*How much value does <name>’s work add to society?*”, $r = .93$. The items are similar to those used in the literature (e.g., Davis, 1989; Elliott, 2007).

Working hard was measured with adapted items from the symbolic racism theory (Henry & Sears, 2002; Sears & Henry, 2003), *“<name> is only doing what she is doing because she is lazy*” (recoded) and *“<name> is probably just working as hard as everyone else*”, $r = .57$.

Symbolic threat was measured with items adapted from Gonzalez et al. (2008), “*The identity of people with normal jobs is being threatened because there are too many <jobs>*” and “*Norms and values are being threatened because of the presence of <jobs>*”, $r = .77$. Because of a copy and paste error, we used the wrong profession for football players and excluded the two items from all analyses involving football players.

Contact was measured with items adapted from Reimer et al. (2021) “*In the past year, how often have you <consumed product>?*” and “*In the past year, how often have you interacted with <people doing job>?*”, $r = .38$.

Openness to new experience was measured with the same items as in Experiment 1 ($\alpha = .75$) (de Vries, 2013), as were attitudes toward digitalization, $r = .79$, and political orientation, $r = .87$.

**3.2. Results**

First, we ran five 2 (condition: Established vs. digital jobs) × 9 (jobs) mixed-ANOVAS, one for the dependent variable (attitudes) and four for all mediators (perceived usefulness, hard-working, symbolic threat, and contact). All main effects and interactions were significant at $p < .001$ (Table S5; Table S6 for descriptive statistics and effect sizes). As predicted, we found that participants hold more positive attitudes toward established jobs than digital jobs, replicating Experiment 1 (Figure 2), perceived established jobs as more useful (Figure S7), perceived people doing established jobs as more hard-working (Figure S8), and less of a symbolic threat (Figure S9). Finally, on average, participants had more contact with established jobs (Figure S10). The strength of the effect depended on the job: We found that people were holding, on average, the least positive attitudes toward models and most positive attitudes toward lecturers (Figure 2), and perceived the two target groups also as least/most useful as well as least/most hard-working (model/lecturer).

Interestingly, the interviewer pair (talk-show host vs. podcaster) emerged as a unique case where the digital version was perceived more favorably than the established counterpart ($M = 4.59$ vs. $M = 4.26$, $p = .002$). This suggests that for certain information-heavy roles, the digital medium (podcasting) may have already achieved a high degree of "normalization" and perceived utility, potentially outweighing symbolic concerns.

**Figure 2**

*Attitudes toward a job depend on the type of job and condition (established vs. digital; Experiment 2). Error bars represent 95%-CIs.*

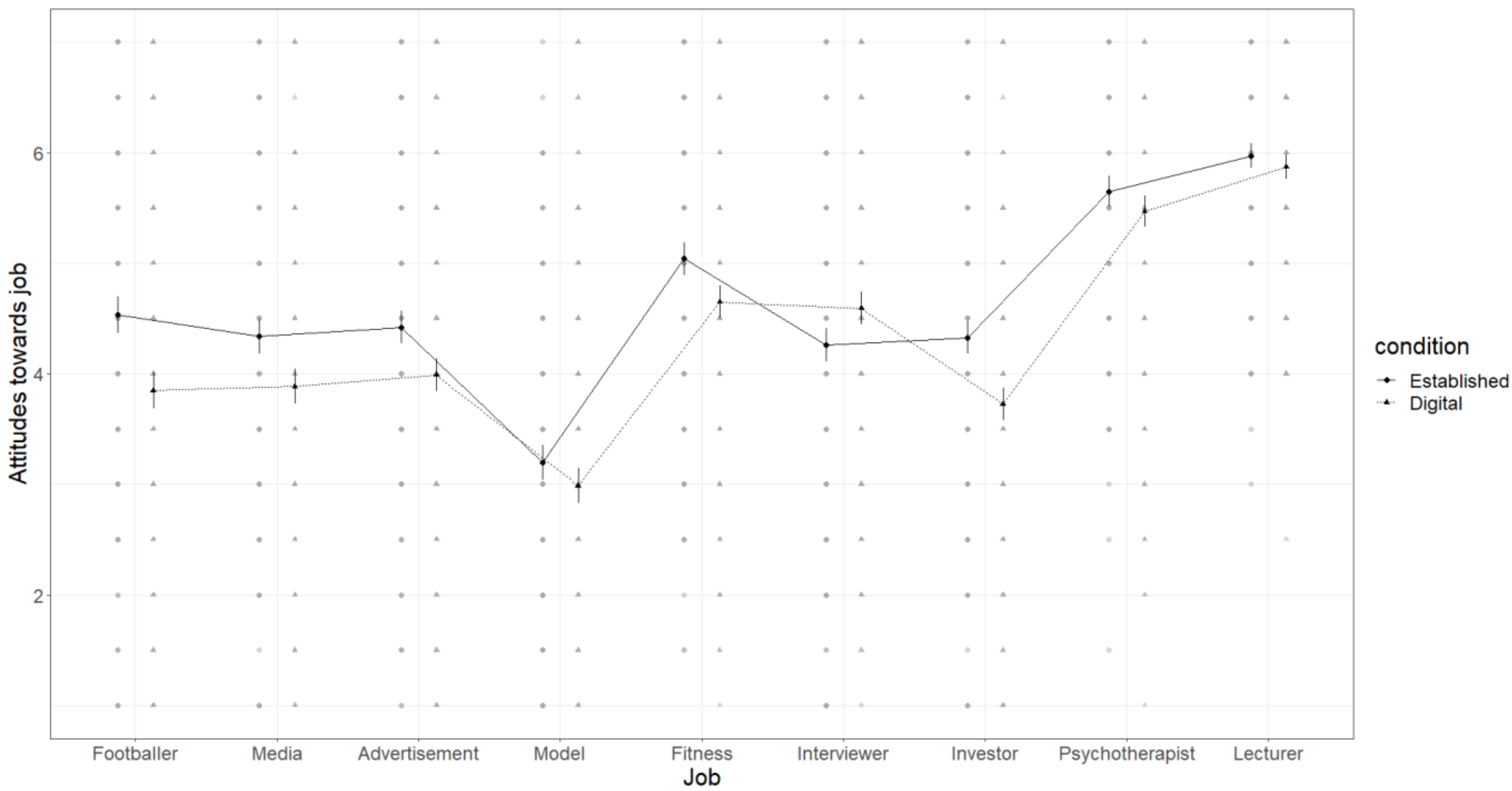

To test whether perceived usefulness, hard-working, symbolic threat, and contact mediated the association between condition and attitudes, we again ran a series of nine mediation analyses. The indirect effect was significant (i.e., the 95%-CI did not include zero) for all professions except lecturer (Table S7). Exploratory follow-up analyses revealed that all four mediators functioned as partial mediators when entered individually.

To test whether openness to new experiences, attitudes toward digitalization, political orientation, or age moderated any of the effects of condition (digital vs. established jobs), we again ran the same four linear-mixed effects models with condition and one of the four moderators at a time as predictors as well as their interactions (Figures S11-S14; Table S8 for correlation coefficients). However, none of the interactions reached significance, suggesting that openness to new experiences, attitudes toward digitalization, political orientation, and age did not influence in our experiment the effect of condition. Of the four main effects, only attitudes toward digitalization was positively associated with attitudes toward the different jobs, $B = .14$, $SE = .04$, $p = .002$.

Finally, we tested whether any of the mediators, moderators, or demographic variables would uniquely predict attitudes toward jobs separately for each condition (established vs digital jobs), using linear-mixed effects models. Interestingly, usefulness, hard work, and symbolic threat were the strongest predictors of attitudes toward both digital and established jobs, followed by attitudes toward digitalization and openness (Table 2). Also as one might expect, political orientation was negatively associated with attitudes toward digital job, suggesting that left-wingers hold on average more positive views than right-wingers. Surprisingly, age again did not predict attitude toward digital jobs.

**Table 2**

*Predictors of attitudes toward job separately for each condition (Experiment 2)*

| | Digital jobs | | | Established jobs | | |
|---|---|---|---|---|---|---|
| | *B* | *SE* | *p* | *B* | *SE* | *p* |
| Usefulness | 0.50 | .01 | <.001 | 0.53 | .01 | <.001 |
| Contact | 0.03 | .01 | .056 | 0.04 | .01 | .001 |
| Hard work | 0.39 | .03 | <.001 | 0.38 | .03 | <.001 |
| Symbolic threat | -0.10 | .02 | <.001 | -0.13 | .02 | <.001 |
| Openness | -0.01 | .02 | .691 | -0.02 | .02 | .347 |
| Attitudes toward digitalization | 0.04 | .02 | .099 | 0.03 | .02 | .195 |
| Political orientation | -0.01 | .02 | .547 | -0.02 | .02 | .349 |
| Age | -0.00 | .00 | .466 | -0.00 | .00 | .455 |
| Gender (1: men, 2: women) | 0.00 | .06 | .948 | 0.02 | .06 | .714 |
| Education level | 0.00 | .03 | .910 | -0.00 | .03 | .928 |
| Income | 0.00 | .03 | .953 | -0.03 | .03 | .315 |

*Note.* Multicollinearity was not an issue, all VIFs < 2.

## 4. General Discussion

Across two experiments, we consistently found that people hold more negative attitudes toward people working in digital jobs compared to people working in more established jobs. Interestingly, these effects were consistent across participants' openness to new experiences, attitudes toward digitalization, political orientation, and age. Contact, perceived usefulness, the perception that they are hard-working, and symbolic threat partially mediated the effect of job condition and attitudes, thereby providing evidence for underlying mechanisms: Digital jobs are perceived as more negative partly because they are seen as less useful, people doing them are perceived as less hard-working and more of a threat to societal values, and because people have less contact with individuals in digital professions. The consistent, independent contribution of perceived usefulness shows how theories from information systems can enhance our understanding of how digital jobs are perceived, beyond what social psychological theories alone can explain.

### 4.1. Theoretical contributions

Perceived usefulness was originally defined in relation to technology rather than jobs or people (Davis, 1989). Our research expands this concept, demonstrating its relevance in predicting attitudes toward individuals in specific jobs. Our findings suggest that the contribution of a person's work to society also shapes attitudes toward this person. For instance, e-Sports players are sometimes perceived as too lazy to engage in 'real work' (BBC Radio 4, 2019), leading to a diminished perception of their societal contributions. We speculate that this association stems, at least in part, from our participants' assumption that individuals freely choose their jobs. Otherwise, their work would not necessarily reflect their personal interests and values. This reasoning aligns with previous research showing that people are judged differently depending on whether they were considered responsible for an action (Coelho et al., 2023; Schellenberg & Bem, 1998). The effects of perceived usefulness on job attitudes provide novel support for the 'social utility-based acceptance/rejection model' (Dambrun, 2024), which predicts that attitudes toward individuals are shaped, at least in part, by their perceived societal contributions (i.e., their perceived usefulness).

Our results further suggest that digital jobs are partly perceived more negatively because they are considered a threat to the values of traditional job holders. These findings resonate with the integrated threat theory, which postulates that prejudice arises from perceived symbolic and realistic threats (Stephan et al., 1999; Stephan & Stephan, 2000). The symbolic threat could arise from rooted social norms and structures that commonly value certain skills and roles more than others. For example, some more established jobs might have traditionally been seen as more "valuable" (van Tilburg et al., 2023), while digital jobs, which may require a different set of skills, might have been perceived as less important. This can create a perception that individuals in digital jobs are less skilled than their traditional counterparts.

Realistic threats, conversely, could arise from the rapid evolution of technology and digitalization. As digital professions or platforms become more present in our day-to-day lives, they could disrupt traditional industries, leading to job loss or devaluing certain skills. Such an impact can induce fear and resistance among those in traditional jobs, who may feel threatened by the uncertain economic future brought by digitalization. Our findings warrant further exploration. Future research could provide more insights by investigating how different digital jobs are perceived as threatening. It could also explore how these perceptions vary among job holders, which would provide a more nuanced understanding of the attitudes.

Our research further supports contact theory (Allport, 1958): Less frequent contact with specific jobs was associated with fewer favorable views. We believe this was primarily driven by a lack of understanding of digital jobs, potentially mostly by older adults. When interacting with tools such as tablets, commonly used to engage with the new jobs described in this study (e.g., following digital influencer), older adults reported facing barriers such as inadequacy, concern about the complexity of technologies, or even lack of knowledge and confidence (Vaportzis et al., 2017). Consequently, this might lead them not to engage with these new job types and have a lower understanding of how they work. Future research might want to test how much knowledge people have about specific jobs, which in turn can be compared with the level of contact in predicting attitudes. People might have various misconceptions about digital jobs, such as not paying well or not requiring significant effort and time. While we indicated for each job pays enough for a good life (Table S1), people might still interpret this differently. For example, participants might assume that professional e-football players earn less than traditional professional football players, even though only a small number of traditional football players earn millions at the top level.

Two findings, which are either inconsistent or non-significant, merit attention: Age was not associated with attitudes toward digital jobs in either experiment, while political orientation showed an association only in Experiment 1. This challenges the stereotype that older individuals are significantly more techno-skeptical than younger ones (e.g., Chung et al., 2010). However, meta-analytic evidence suggests a negative, albeit weak, association between age and technology acceptance (Hauk et al., 2018). More research is needed to understand whether age differences in technology acceptance have diminished over recent years – possibly because older people are now more familiar with technology (cf. Staddon, 2020). Alternatively, this finding could be unique to our study design, or it may reflect a limitation in our non-representative sample if we recruited a more tech-savvy group of older participants. Regarding political orientation, although conservative people attach more importance to tradition and preservation of the status quo (Caprara et al., 2006; Schwartz, 1992), they are also more likely to support free markets (Hunter & Milofsky, 2007) and have stronger beliefs in free will (Everett et al., 2021). Theoretically, belief in free markets and free will should be associated with more positive attitudes toward newly invented jobs. Our findings suggest that these opposing motivations might counteract each other, resulting in only inconsistent associations with attitudes toward digital jobs.

**4.2. Implications**

In the context of our rapidly digitalizing job market, the implications of our findings could be particularly useful. The perception of new and digital jobs can have societal implications, as greater acceptance can lead to greater equality given that they require fewer resources and smaller social networks to be launched. The study of digital job adoption and acceptance is itself relatively new and underexplored, and our findings contribute a more nuanced framework for understanding it. A better understanding of digital jobs, potentially through educational campaigns or firsthand experience, may help develop positive attitudes toward them (Celuch et al., 2022). This point is particularly crucial because criticism has been shown to reduce the online engagement levels of those who are criticized (Urbaniak et al., 2022), making contact less likely and thereby potentially leading to a downwards spiral. This perception can also discourage the creation of new digital jobs, as potential entrepreneurs may hesitate to launch digital-focused startups because of concerns about societal judgment.

Improving attitudes toward people working in new digital jobs can also benefit their mental health by reducing the stigma associated with their work (cf. Livingston et al., 2012). Research consistently shows that stigmatization can negatively impact mental health (Meadows & Bombak, 2019; Thornicroft et al., 2022). There is some preliminary evidence that stigma might also impact people who are working in new digital jobs: E-sports players are more likely to report worse mental health (Soares et al., 2022). Therefore, improving attitudes toward people in new digital jobs can enhance their mental health by reducing stigma. This, in turn, can have implications for digital career development, as less stigma helps individuals pursue their own goals (Corrigan, 1998). Relatedly, improving attitudes will make it easier to integrate digital jobs into the mainstream workforce. Research has established numerous methods to improve attitudes towards other people (e.g., immigrants, people with different political views, minorities), such as contact or emphasizing similarities in psychological variables that also work well online (Hanel et al., 2019; Pettigrew & Tropp, 2006; Voelkel et al., 2024). Future research can test which of these methods works best for improving attitudes towards people with a digital job. Improving attitudes is necessary given the projected rise of digital jobs in the future (World Economic Forum, 2024). Moreover, policymakers can benefit from our findings by developing targeted interventions aiming to reduce stigma and promote a more positive perception of these digital careers. Such policies can lead to a more inclusive job market and better support for digital professionals. Overall, our findings have important implications for digital career development and policy-making.

Our findings offer "implementable levers" for organizations and platforms in the digital economy. First, to combat the "uselessness" stigma, digital platforms should encourage creators to explicitly communicate the societal value or educational aspect of their content – moving the narrative from "lifestyle" to "contribution." Second, addressing symbolic threat requires framing digital work in terms of professional standards and discipline, aligning it with traditional values of "hard work." Finally, organizations can foster acceptance by increasing structured contact; for example, by integrating digital content creators into traditional marketing teams, thereby demystifying their daily routines and skills (Vivaldini & de Sousa, 2024).

**4.3. Limitations**

One limitation of our project is that some of the established and the digital jobs might not have been perfectly matched, even though we attempted to align digital and more established jobs as closely as possible. For example, a professional e-football player presumably requires less extensive physical training than a professional football player. However, we argue that other professions, such as advertisers or investors, are more comparable also in terms of income and education level required, and we find the effects for them as well. Further, we acknowledge that the boundaries between the established and digital versions of the jobs we selected are somewhat blurred (e.g., talk shows are often also broadcast on the internet). When referring to established versus digital jobs, we intend to indicate relatively more established versus relatively more digital positions, rather than implying a strict dichotomy.

One might argue that digital jobs are indeed less useful. We believe that there is some variance regarding the usefulness of digital jobs, as there is for more established jobs. For instance, roughly 1 in 4 people working in the UK and USA do not find their job meaningful (Ballard, 2021; Nolsoe, 2020). There is also substantial variance in how boring various jobs are perceived, both for jobs that are heavily digitalized but also jobs that are mostly done offline (van Tilburg et al., 2023). Together, we believe it is difficult to objectively determine whether digital jobs are less useful than their more established counterparts. A systematic investigation into the usefulness from economic, environmental, and societal perspectives is difficult. For example, the choice of dependent variables might be influenced by personal biases from the experts (Duarte et al., 2015), which can impact the assessment of job usefulness.

Furthermore, while we selected jobs requiring different skill sets to make our findings more generalizable, there are still some types of jobs we neglected. For instance, future research could investigate whether the findings are reversed for illegal jobs. For example, a criminal hacker might be perceived as less negative than a burglar because burglars tend to invade the personal physical space of a person. In contrast, a hacker 'only' accesses the personal digital space, which might be perceived as less threatening.

Moreover, our participants were all living in one Western country, the UK. While it is common practice to recruit participants from one (Western) country (Thalmayer et al., 2021) and some findings from Western countries replicate in many other countries, including less developed countries (Cohn et al., 2019; Schwartz & Bardi, 2001), we do not know whether our findings will replicate in other countries. We suspect that effects would be less strong in populations who are not using the internet, simply because they might not have heard of many of the new digital jobs. However, we included a range of personality and attitudinal variables such as openness to new experiences, attitudes toward digitalization, and political orientation, and found that our effects were consistent across them (i.e., no interactions between the moderators and condition). Since personality and attitudinal variables tend to vary more within-countries than between-countries (Saucier et al., 2015), we suspect that future studies would overall replicate our effects across other countries.

While our findings demonstrate the value of combining theories from different disciplines to gain a better understanding of attitudes towards jobs by enhancing the predictive power of our statistical models, it is possible that incorporating additional elements from other theories and models could further strengthen this predictive power. For example, building on relevance theory (Schulz, 2003), we speculate that perceiving a job as personally relevant may be associated with more positive attitudes toward that job. Further, perceiving a job as having desirable job characteristics as outlined by the job characteristics theory, such as autonomy, skill variety, and job significance (Hackman & Oldham, 1980), might also be associated with more positive attitudes towards the job and the person doing the job because these characteristics are considered positive (Hackman & Oldham, 1976; Ryan & Deci, 2000).

Additionally, we acknowledge potential confounds in our vignette cues. While we explicitly stated that all individuals earned enough for a "good life," participants may have held pre-existing stereotypes regarding the "ease" of digital work or the formal training required. For instance, the perception that a traditional footballer undergoes more "rigorous physical training" than an e-sports player could influence attitudes beyond the digital/traditional distinction itself. Future studies could use manipulation checks to assess whether perceived effort and training duration are balanced across vignettes.

Future research should employ design remedies such as Delphi-based expert matching for vignettes and preregistered comparability thresholds. Additionally, field experiments in ecosystems where digital entrepreneurship surged post-COVID (e.g., MSMEs) could manipulate usefulness narratives to test if highlighting instrumental value can overcome symbolic stigma (Sreenivasan et al., 2023). Another open question is under what conditions perceived usefulness versus symbolic threat is the stronger driver of attitudes toward digital jobs. Addressing this would require designs that experimentally manipulate the visibility of societal contribution and value alignment independently, which we leave to future work.

### 4.4. Conclusion

Theoretically, our research highlights the advantages of interdisciplinary work by showing how combining theories from different domains can yield novel insights. While psychological theories such as the theory of planned behavior (Ajzen, 1991) initially contributed to shaping information systems theories (Dwivedi et al., 2019), we showed that theories from information systems can now also enrich psychological theories.

## Online Supplemental Materials

### Experiment 1

**Table S1**

*Job vignettes created for Experiment 1*

| Job title | Established job | Digital job |
|---|---|---|
| Footballer | Charlie has a solid career as a football player. Charlie plays for a famous professional team, which competes in championships, and plays against some of the best football players worldwide. Charlie's monthly income allows him to lead a good life. | Charlie has a solid career as an e-sports football player, that is playing football on video game consoles such as Playstation and X-Box. Charlie plays for a famous professional football e-sports team, which competes in online championships, and plays against some of the best e-sports players worldwide. Charlie's monthly incomes allow him to lead a good life. |
| Presenter (Media) | Taylor has a solid career as a TV Presenter. More specifically, Taylor has a famous daily tv show on a big news channel, which discusses various topics, from politics to gossip about famous artists. Taylor's income allows her to lead a good life. | Taylor has a solid career as a YouTuber. More specifically, Taylor has a famous channel on YouTube, with daily videos discussing a variety of topics, from politics to gossip about famous artists. Taylor's income allows her to have a good life. |
| Advertiser | Anne has a solid career in advertising. Anne creates ads for various products that are broadcasted on TV and printed in magazines. Anne's salary depends on the number of advertisements she develops and the success thereof. Overall, Anne's income allows her to have a good life. | Anne has a solid career as a digital influencer. Anne creates ads for various products that she posts on social media accounts. Anne's salary depends on the number of advertisements she develops and the success thereof. Overall, Anne's income allows her to have a good life. |
| Adult model | Sasha has a solid career as a model in adult magazines. Sasha regularly participates in erotic photo shootings with professional photographers booked by adult magazines. | Sasha has a solid career as an OnlyFans model, a website on which content creators can sell pictures and videos of themselves. Sasha regularly uploads homemade erotic photos. While Sasha's monthly |
| | While Sasha's monthly income can vary, it overall allows her to have a good life. | income can vary, it overall allows her to have a good life. |
| Fitness instructor | Tom has a solid career as a personal trainer. Tom provides an in-person service to develop a personalized workout plan for his clients. Tom's salary depends on the number of clients per month. Overall, Tom's income allows him to lead a good life. | Tom has a solid career in providing workout tips on social media. Tom posts videos online showing various exercises that can be performed safely. Tom's salary depends on the number of views his social media posts receive. Overall, Tom's income allows him to lead a good life. |
| Interviewer | Paul has a solid career as a talk-show host: He has a famous TV show on a large news channel, and interviews celebrities from different areas, including actors, politicians, artists, and scientists. Paul's income allows him to have a good life. | Paul has a solid career as a podcaster: He has integrated channels on multiple platforms (e.g., Spotify, Deezer, Youtube), and interviews celebrities from different areas, including actors, politicians, artists, and scientists. He then shares the recorded interviews on the platforms. Paul's income allows him to have a good life. |
| Investor | Jenny has a solid career as a stock market investor. Jenny buys and sells shares based on her market analyses. Jenny tries to minimize risks and maximize returns. While Jenny's monthly income varies, it allows her to have a good life. | Jenny has a solid career as a bitcoin investor (bitcoins are a decentralized digital crypto-currency that can be bought and sold and which values can increase or decrease depending on supply and demand). She buys and sells bitcoins based on her analyses of the cryptocurrency's fluctuations. Jenny tries to minimize risks and maximize returns. While Jenny's monthly income varies, it allows her to have a good life. |
| Psychotherapist | Cate has a solid career as a psychotherapist. Cate treats her patients in her office. Cate's salary depends on the number of clients she treats. Overall, Cate's income allows her to have a good life. | Cate has a solid career as an online psychotherapist. Cate treats her patients through platforms such as Zoom and Microsoft Teams. Cate's salary depends on the number of clients she treats. Overall, Cate's income allows her to have a good life. |
| Lecturer | Gabriel has a solid career as a lecturer. Gabriel works for a campus university, and the role typically includes delivering in-person classes and providing student support. Gabriel’s income allows him to lead a good life. | Gabriel has a solid career as a lecturer. Gabriel works for an online-only university, and the role typically includes delivering remote classes and providing student support through platforms such as Zoom and Microsoft Teams. Gabriel’s income allows him to lead a good life. |

**Figure S1**

*Experiment 1: Perceived usefulness of a job depends on the type of job and condition (established vs digital). Error bars represent 95%-CIs.*

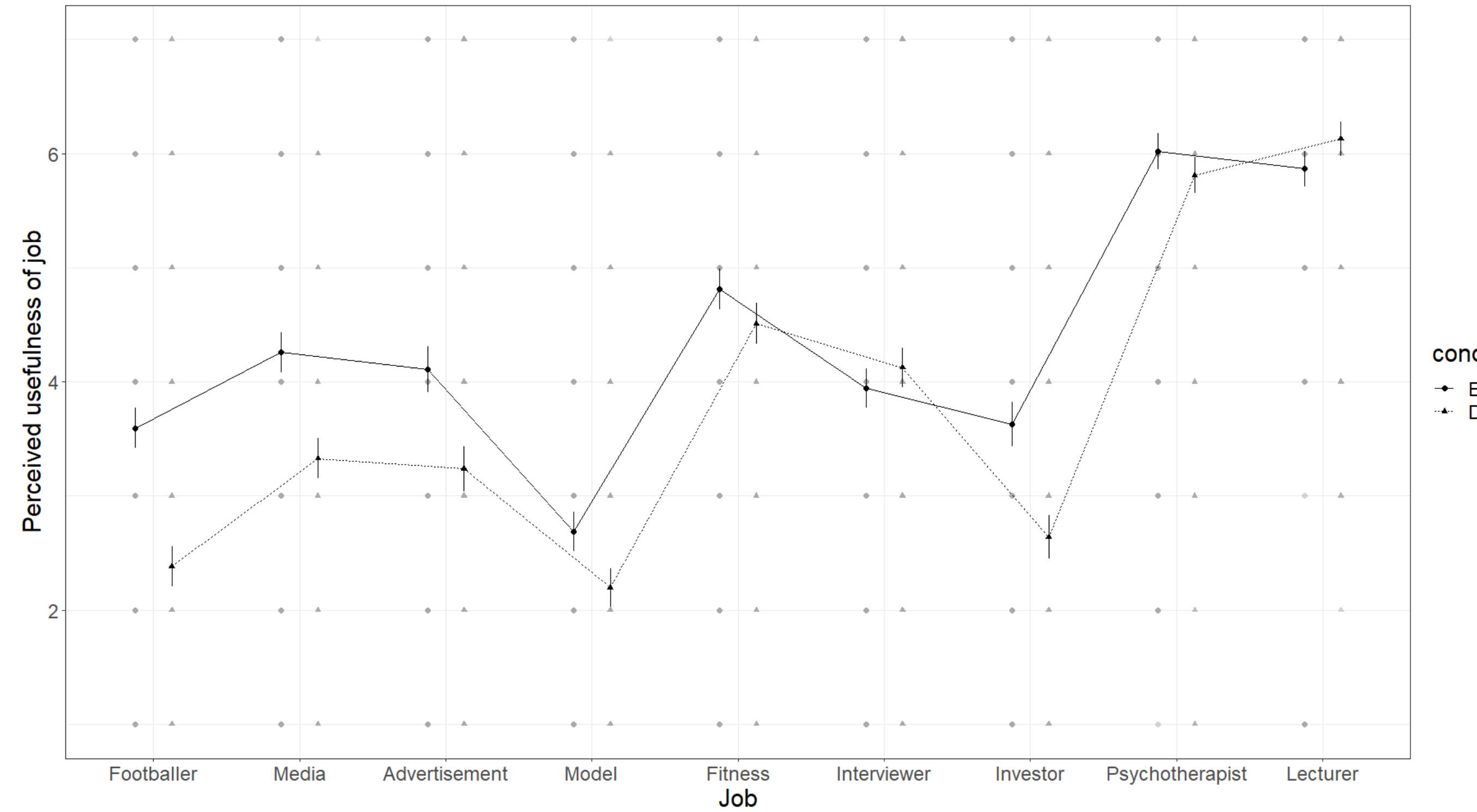

**Figure S2**

*Experiment 1: Amount of contact depends on the type of job and condition (established vs digital). Error bars represent 95%-CIs.*

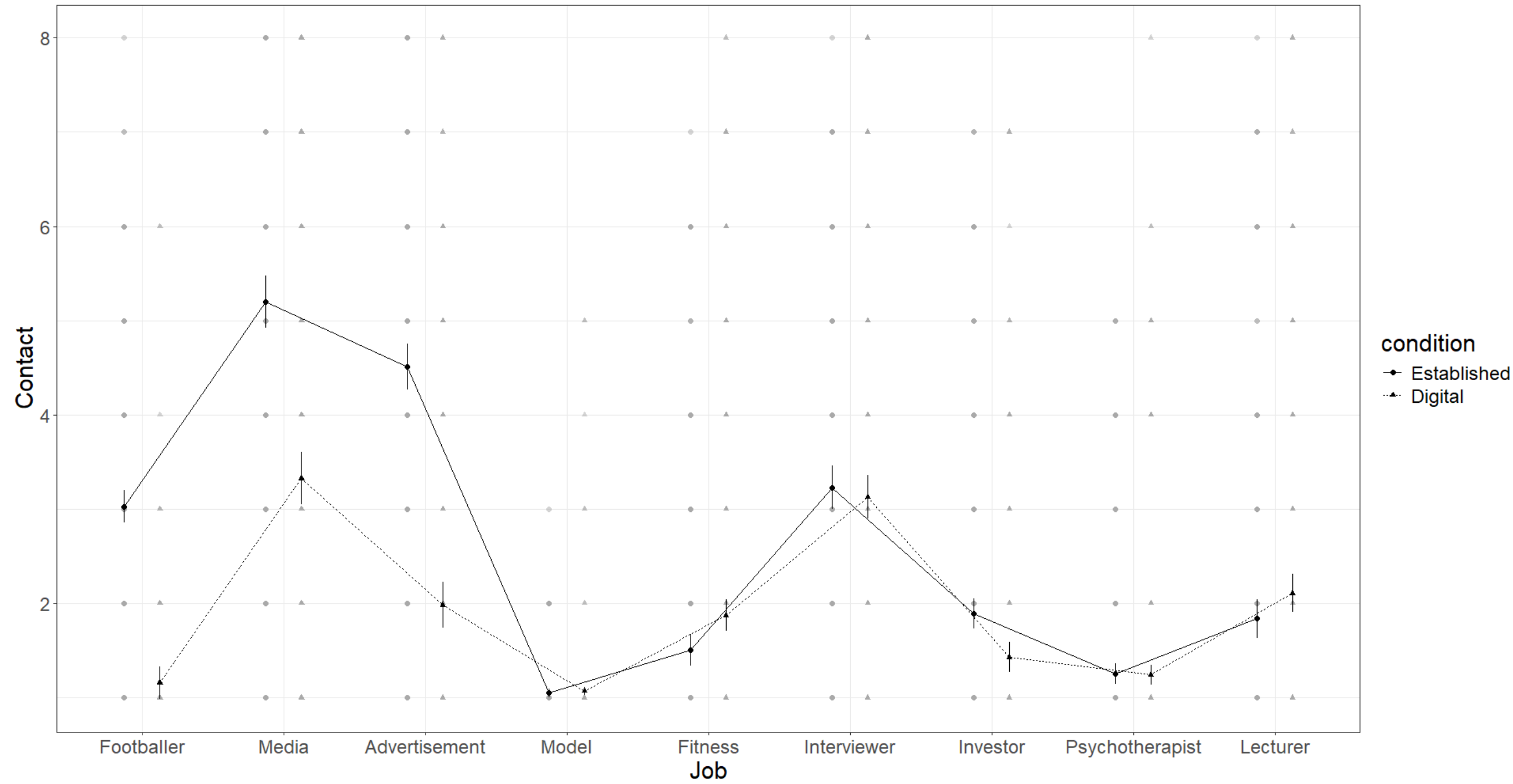

**Figure S3**

*Experiment 1: Association between openness and attitudes towards jobs, separated by condition and job type*

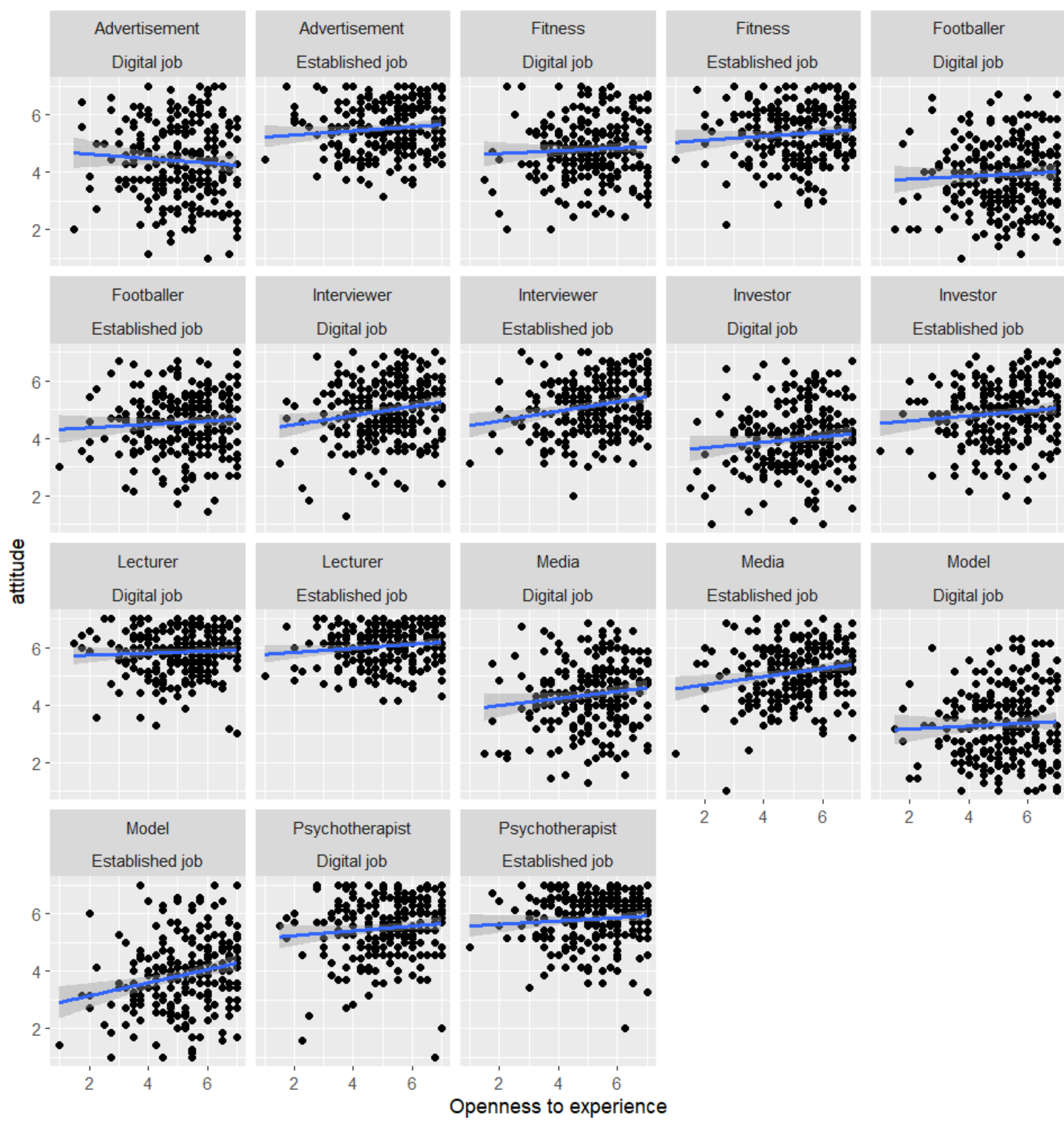

**Figure S4**

*Experiment 1: Association between attitudes towards digitalisation and attitudes towards jobs, separated by condition and job type*

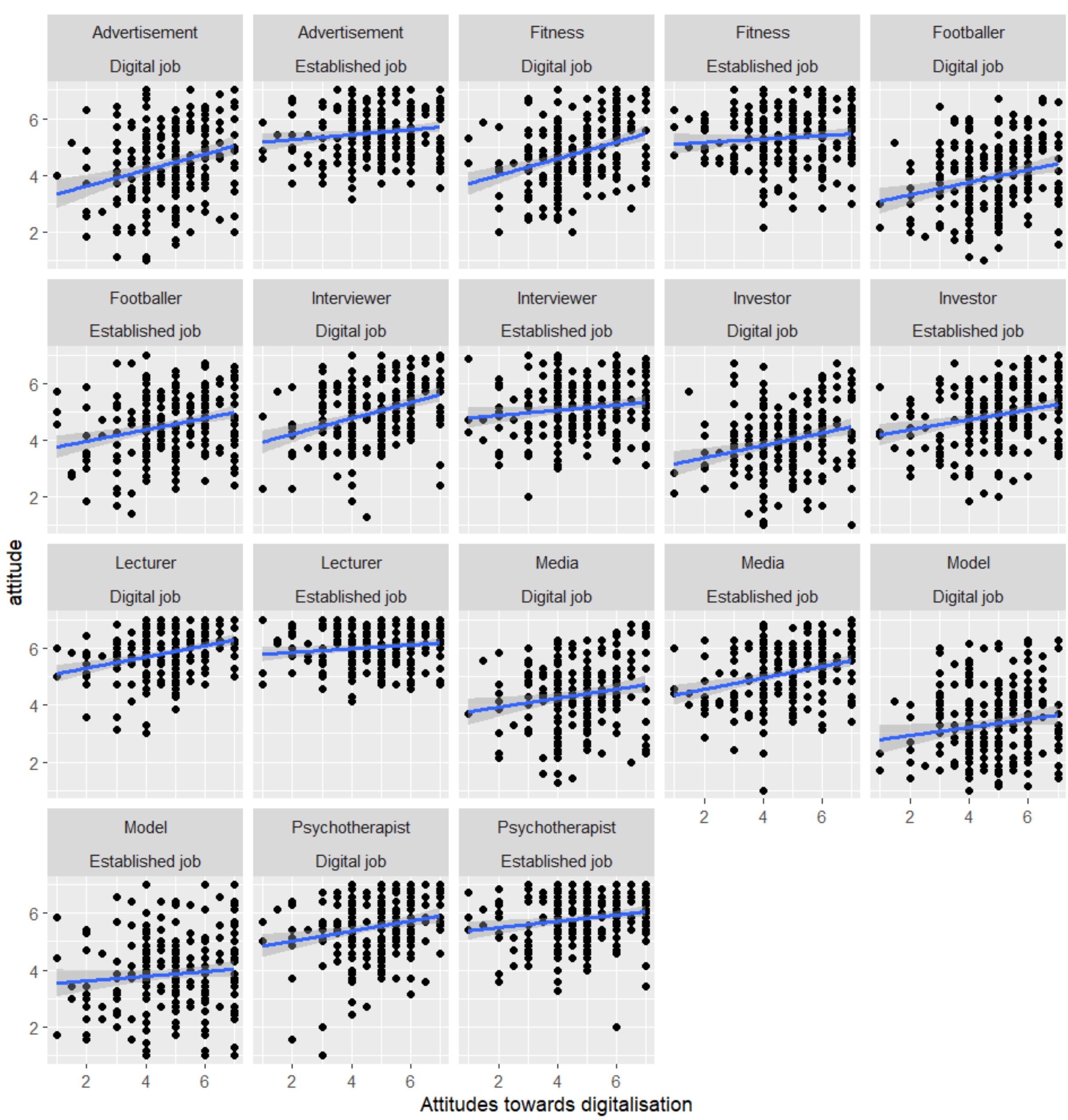

**Figure S5**

*Experiment 1: Association between political orientation and attitudes towards jobs, separated by condition and job type*

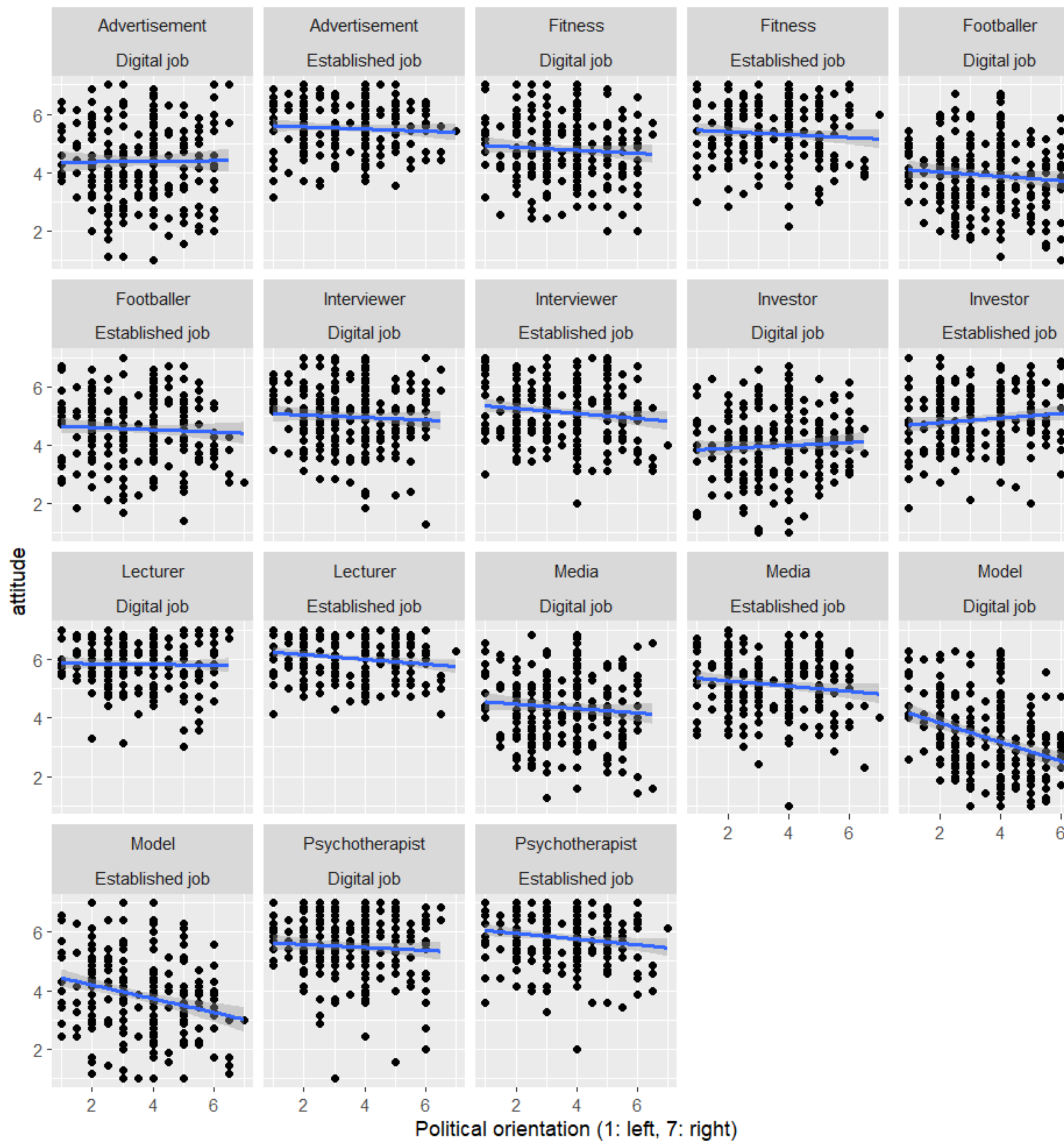

**Figure S6**

*Experiment 1: Association between age and attitudes towards jobs, separated by condition and job type*

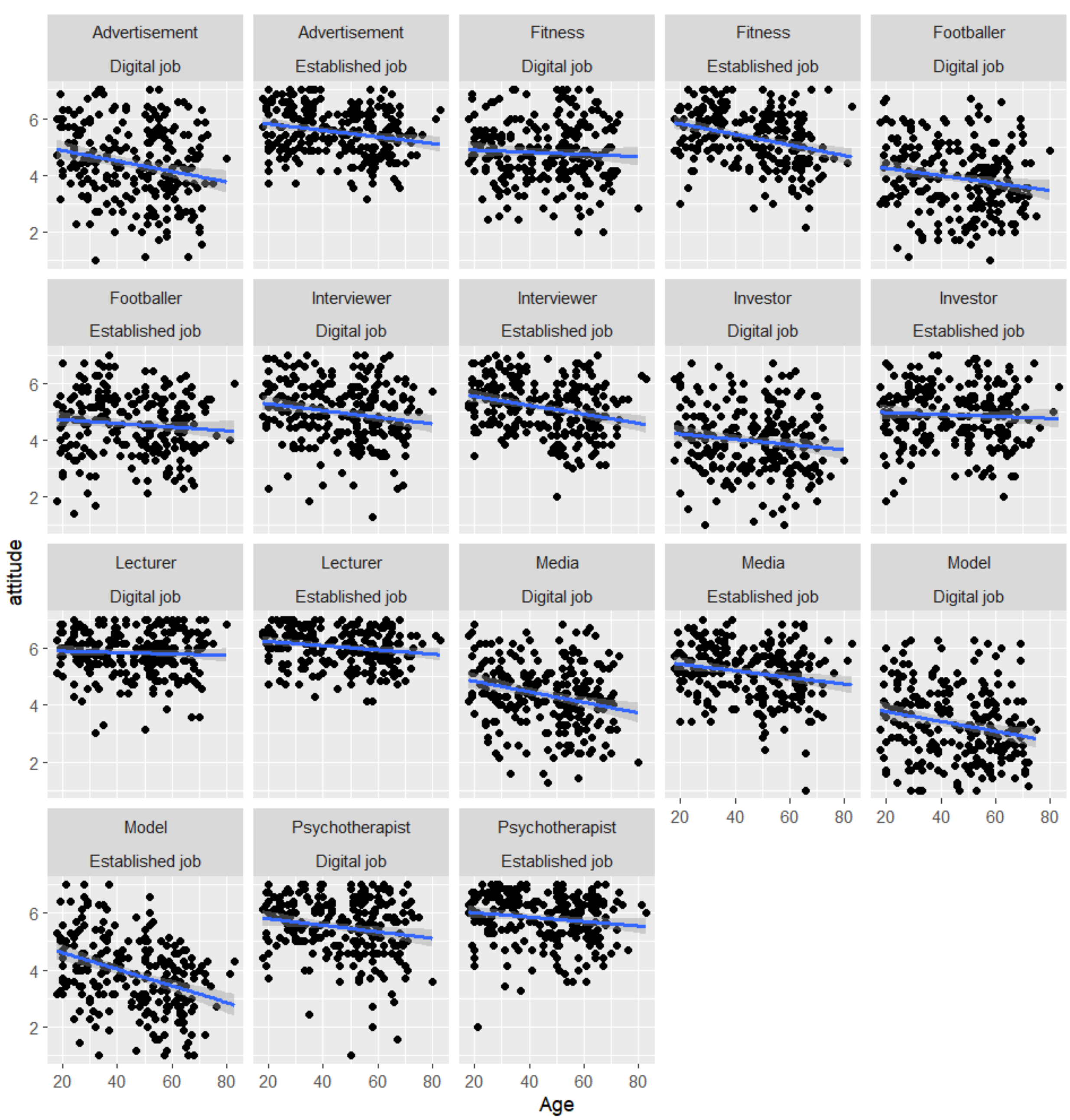

**Table S2**

*Experiment 1: Mean comparisons of condition (established vs digital) separated by job type and dependent variable*

| | | Established | | Digital | | | | | |
|---|---|---|---|---|---|---|---|---|---|
| DV | Job | *M* | *SD* | *M* | *SD* | *t-value* | *p-value* | *d* | *U3* |
| attitude | Advertisement | 5.51 | 0.85 | 4.42 | 1.34 | 10.93 | <.0001 | 0.98 | 83.53 |
| attitude | Fitness | 5.32 | 0.97 | 4.81 | 1.07 | 5.62 | <.0001 | 0.5 | 69.23 |
| attitude | Footballer | 4.58 | 1.1 | 3.91 | 1.17 | 6.61 | <.0001 | 0.59 | 72.24 |
| attitude | Interviewer | 5.14 | 0.96 | 4.98 | 1.06 | 1.7 | .0904 | 0.15 | 56.02 |
| attitude | Investor | 4.89 | 1.03 | 4 | 1.14 | 9.25 | <.0001 | 0.83 | 79.55 |
| attitude | Lecturer | 6.05 | 0.67 | 5.85 | 0.78 | 3.06 | .0023 | 0.27 | 60.77 |
| attitude | Media | 5.14 | 0.96 | 4.39 | 1.19 | 7.8 | <.0001 | 0.7 | 75.7 |
| attitude | Model | 3.88 | 1.27 | 3.33 | 1.28 | 4.78 | <.0001 | 0.43 | 66.51 |
| attitude | Psychotherapist | 5.83 | 0.86 | 5.51 | 1.02 | 3.86 | <.0001 | 0.34 | 63.49 |
| contact | Advertisement | 4.51 | 2.17 | 1.98 | 1.71 | 14.56 | <.0001 | 1.3 | 90.31 |
| contact | Fitness | 1.5 | 1.13 | 1.87 | 1.54 | -3.05 | .0024 | -0.27 | 39.26 |
| contact | Footballer | 3.03 | 1.85 | 1.16 | 0.58 | 15.31 | <.0001 | 1.37 | 91.41 |
| contact | Interviewer | 3.23 | 1.66 | 3.13 | 2.06 | 0.6 | .5477 | 0.05 | 52.14 |
| contact | Investor | 1.88 | 1.46 | 1.43 | 1.1 | 3.97 | <.0001 | 0.35 | 63.84 |
| contact | Lecturer | 1.84 | 1.6 | 2.11 | 1.69 | -1.79 | .0746 | -0.16 | 43.66 |
| contact | Media | 5.2 | 1.96 | 3.32 | 2.47 | 9.43 | <.0001 | 0.84 | 80.01 |
| contact | Model | 1.05 | 0.23 | 1.07 | 0.45 | -0.63 | .5308 | -0.06 | 47.77 |
| contact | Psychotherapist | 1.25 | 0.77 | 1.24 | 0.92 | 0.11 | .9162 | 0.01 | 50.37 |
| usefulness | Advertisement | 4.12 | 1.49 | 3.24 | 1.69 | 6.16 | <.0001 | 0.55 | 70.87 |
| usefulness | Fitness | 4.81 | 1.44 | 4.51 | 1.41 | 2.34 | .0198 | 0.21 | 58.27 |
| usefulness | Footballer | 3.61 | 1.45 | 2.38 | 1.36 | 9.73 | <.0001 | 0.87 | 80.75 |
| usefulness | Interviewer | 3.94 | 1.45 | 4.12 | 1.33 | -1.44 | .1496 | -0.13 | 44.87 |
| usefulness | Investor | 3.64 | 1.63 | 2.64 | 1.44 | 7.27 | <.0001 | 0.65 | 74.19 |
| usefulness | Lecturer | 5.87 | 1.44 | 6.13 | 1 | -2.31 | .0212 | -0.21 | 41.81 |
| usefulness | Media | 4.26 | 1.45 | 3.33 | 1.4 | 7.31 | <.0001 | 0.65 | 74.31 |
| usefulness | Model | 2.69 | 1.4 | 2.2 | 1.33 | 4.08 | <.0001 | 0.36 | 64.22 |
| usefulness | Psychotherapist | 6.02 | 1.28 | 5.81 | 1.25 | 1.83 | .0676 | 0.16 | 56.49 |

*Note*. d: Cohen's d (standardized mean difference). U3: Cohen's U3 (percentage of participants who hold more favourable attitudes towards the established job compared to the average of the digital job). We also report Cohen's U3 because it is perceived as more informative than several other effect sizes (Hanel & Mehler, 2019). All p-values are uncorrected but two-tailed.

**Table S3**

*Experiment 1: Mediation analyses of condition on attitudes with usefulness and contact as mediator (bootstrapped 95%-CIs)*

| | c: total effect | c': Direct effect | Indirect effect [95%-CI] | $R^2$ |
|---|---|---|---|---|
| Advertisement | -1.10*** | -.52*** | -.58 [-.75, -.42] | .59*** |
| Fitness | -.51*** | -.39*** | -.12 [-.26, .01] | .56*** |
| Footballer | -.67*** | .09 | -.76 [-.92, -.61] | .41*** |
| Interviewer | -.15 | -.24*** | .08 [-.04, .20] | .48*** |
| Investor | -.90*** | -.40*** | -.50 [-.64, -.37] | .54*** |
| Lecturer | -.20** | -.28*** | .08 [.01, .14] | .29*** |
| Presenter (media) | -.75*** | -.18* | -.57 [-.72, -.42] | .54*** |
| Model | -.54*** | -.22** | -.32 [-.48, -.17] | .50*** |
| Psychotherapist | -.33*** | -.21*** | -.12 [-.26, .01] | .63*** |

*Note.* Bootstrapped 95%-CIs are based on 5000 resamples. $^*p < .05$, $^{**}p < .01$, $^{***}p < .001$

**Table S4**

*Experiment 1: Correlations between attitudes and the four moderators by job and condition*

| | | Openness | | Digital | | Political | | Age | |
|---|---|---|---|---|---|---|---|---|---|
| job | condition | *r* | *p* | *r* | *p* | *r* | *p* | *r* | *p* |
| Advertisement | Established | .10 | .1050 | .16 | .0108 | -.06 | .3900 | -.21 | .0008 |
| Advertisement | Digital | -.06 | .3240 | .28 | .0000 | .01 | .9370 | -.22 | .0003 |
| Fitness | Established | .11 | .0700 | .11 | .0704 | -.08 | .2270 | -.29 | .0000 |
| Fitness | Digital | .05 | .4360 | .36 | .0000 | -.06 | .3640 | -.05 | .3960 |
| Footballer | Established | .09 | .1540 | .24 | .0002 | -.04 | .5440 | -.09 | .1710 |
| Footballer | Digital | .05 | .4400 | .24 | .0001 | -.09 | .1750 | -.17 | .0062 |
| Interviewer | Established | .24 | .0002 | .16 | .0140 | -.14 | .0317 | -.25 | .0001 |
| Interviewer | Digital | .18 | .0043 | .34 | .0000 | -.05 | .4170 | -.18 | .0048 |
| Investor | Established | .10 | .1330 | .25 | .0001 | .12 | .0675 | -.05 | .4490 |
| Investor | Digital | .11 | .0941 | .25 | .0001 | .06 | .3220 | -.13 | .0337 |
| Lecturer | Established | .13 | .0362 | .15 | .0205 | -.16 | .0108 | -.16 | .0107 |
| Lecturer | Digital | .06 | .3860 | .32 | .0000 | -.02 | .7360 | -.06 | .3730 |
| Media | Established | .18 | .0037 | .30 | .0000 | -.14 | .0239 | -.20 | .0019 |
| Media | Digital | .13 | .0337 | .17 | .0075 | -.06 | .3420 | -.23 | .0002 |
| Model | Established | .22 | .0004 | .06 | .3160 | -.26 | .0000 | -.36 | .0000 |
| Model | Digital | .06 | .3680 | .14 | .0222 | -.34 | .0000 | -.21 | .0009 |
| Psychotherapist | Established | .09 | .1500 | .17 | .0063 | -.17 | .0075 | -.14 | .0304 |
| Psychotherapist | Digital | .11 | .0939 | .22 | .0005 | -.07 | .2960 | -.18 | .0040 |

*Note*. Correlations between attitudes towards the jobs and each of the four moderators. For instance, the correlation between attitudes towards established advertisement job and openness to new experience is $r = .10$, $p = .1050$ (top left of table above). Openness: openness to new experience, digital: attitudes towards digitalization, political: political orientation, age: Age of participants.

## Experiment 2

### Figure S7

*Experiment 2: Perceived usefulness of a job depends on the type of job and condition (established vs digital). Error bars represent 95%-CIs.*

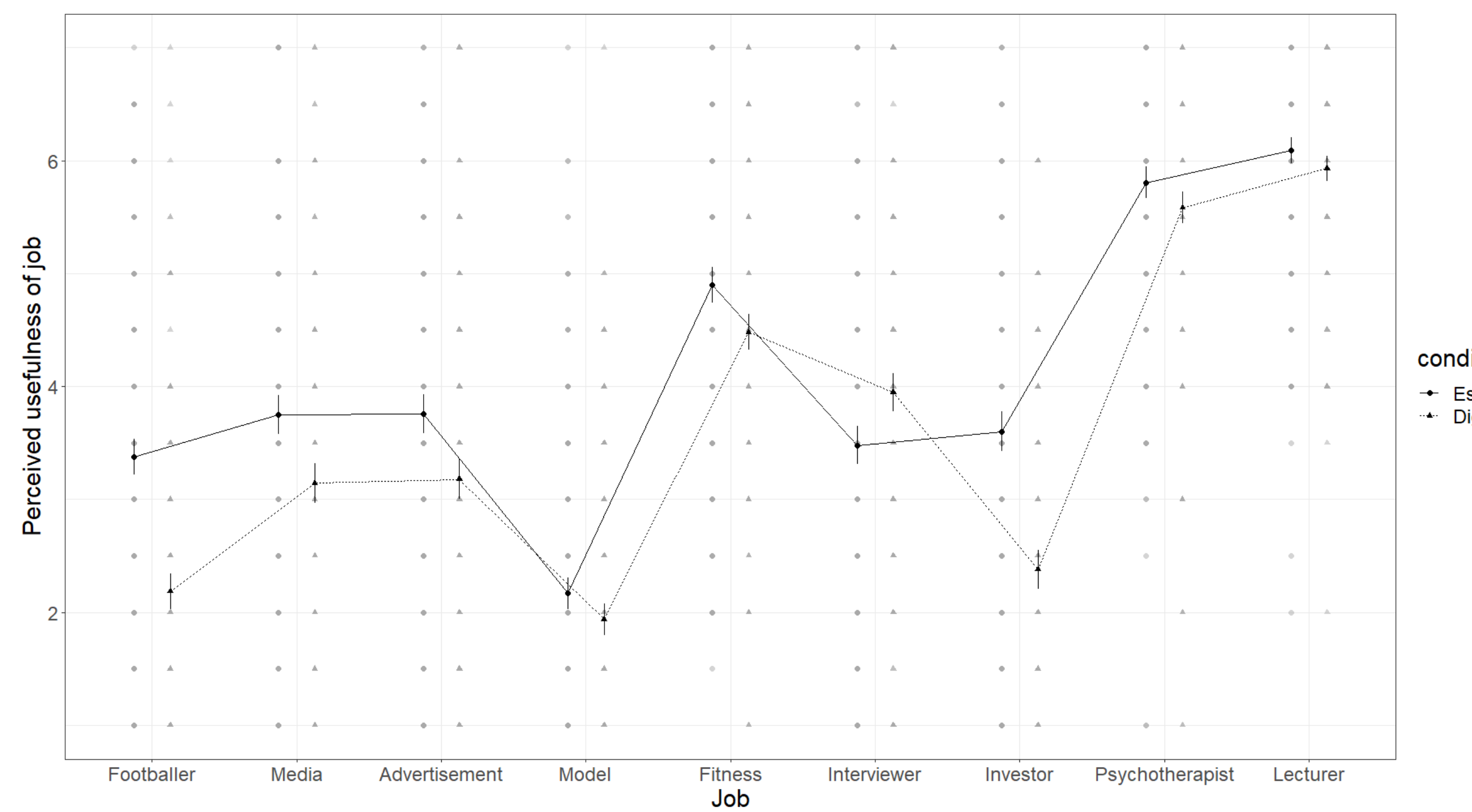

**Figure S8**

*Experiment 2: How hard-working people doing different jobs are perceived depends on the type of job and condition (established vs digital). Error bars represent 95%-CIs.*

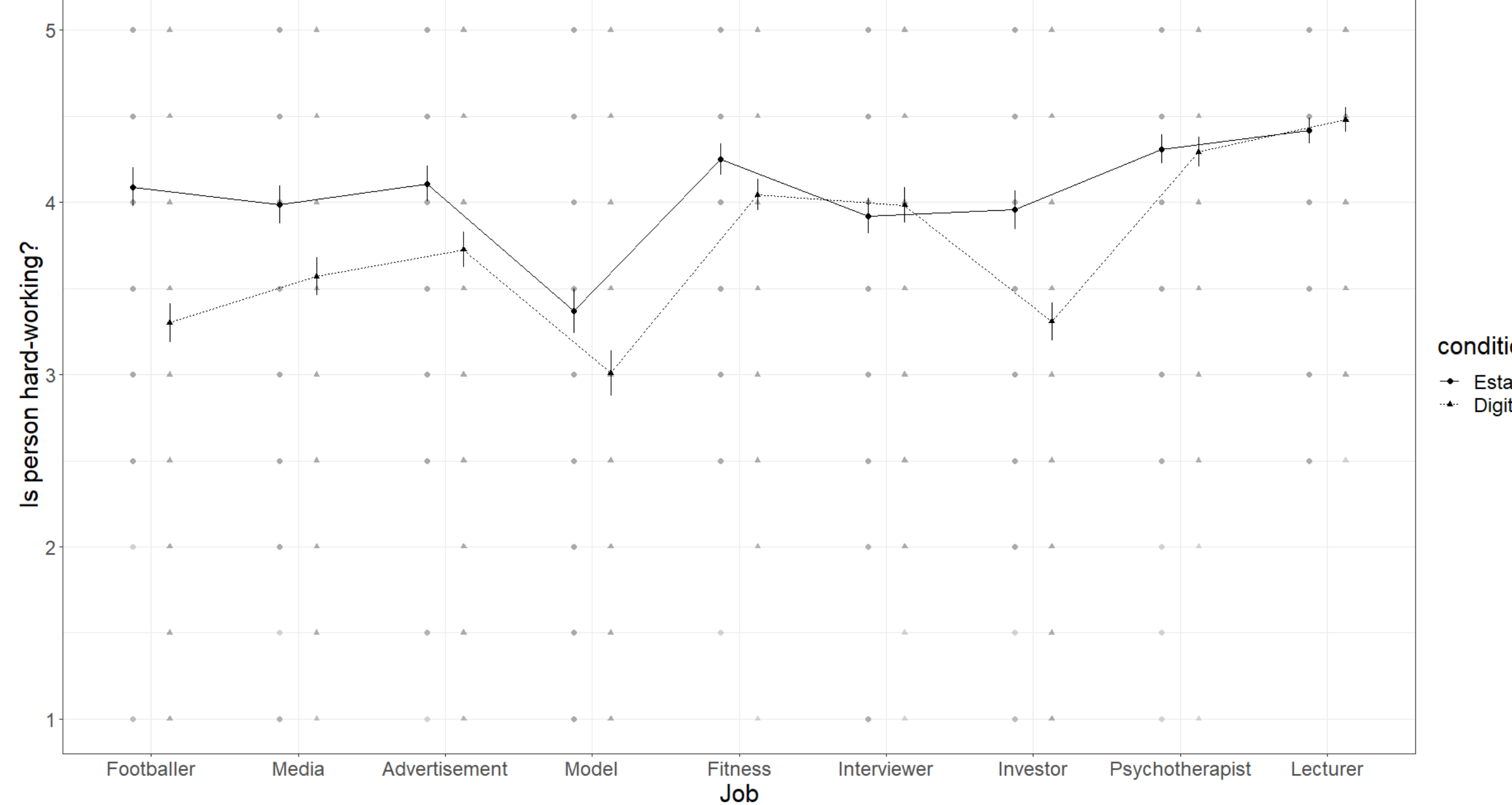

**Figure S9**

*Experiment 2: Symbolic threat of people doing various jobs depends on the type of job and condition (established vs digital). Error bars represent 95%-CIs.*

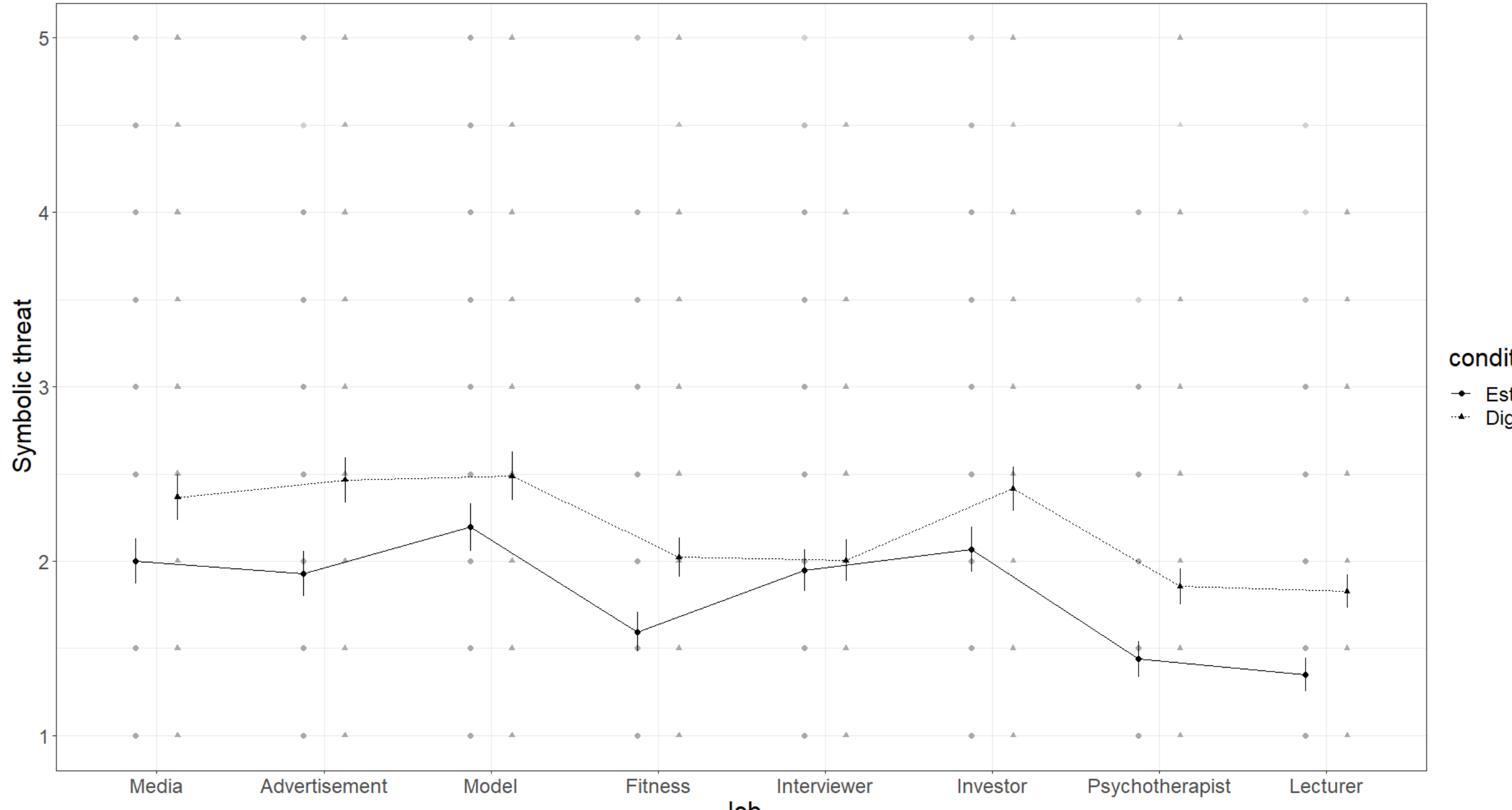

*Note*. Because of a copy and paste error, we used the wrong profession for football players and excluded the two items from all analyses involving football players.

**Figure S10**

*Experiment 2: Contact depends on the type of job and condition (established vs digital). Error bars represent 95%-CIs.*

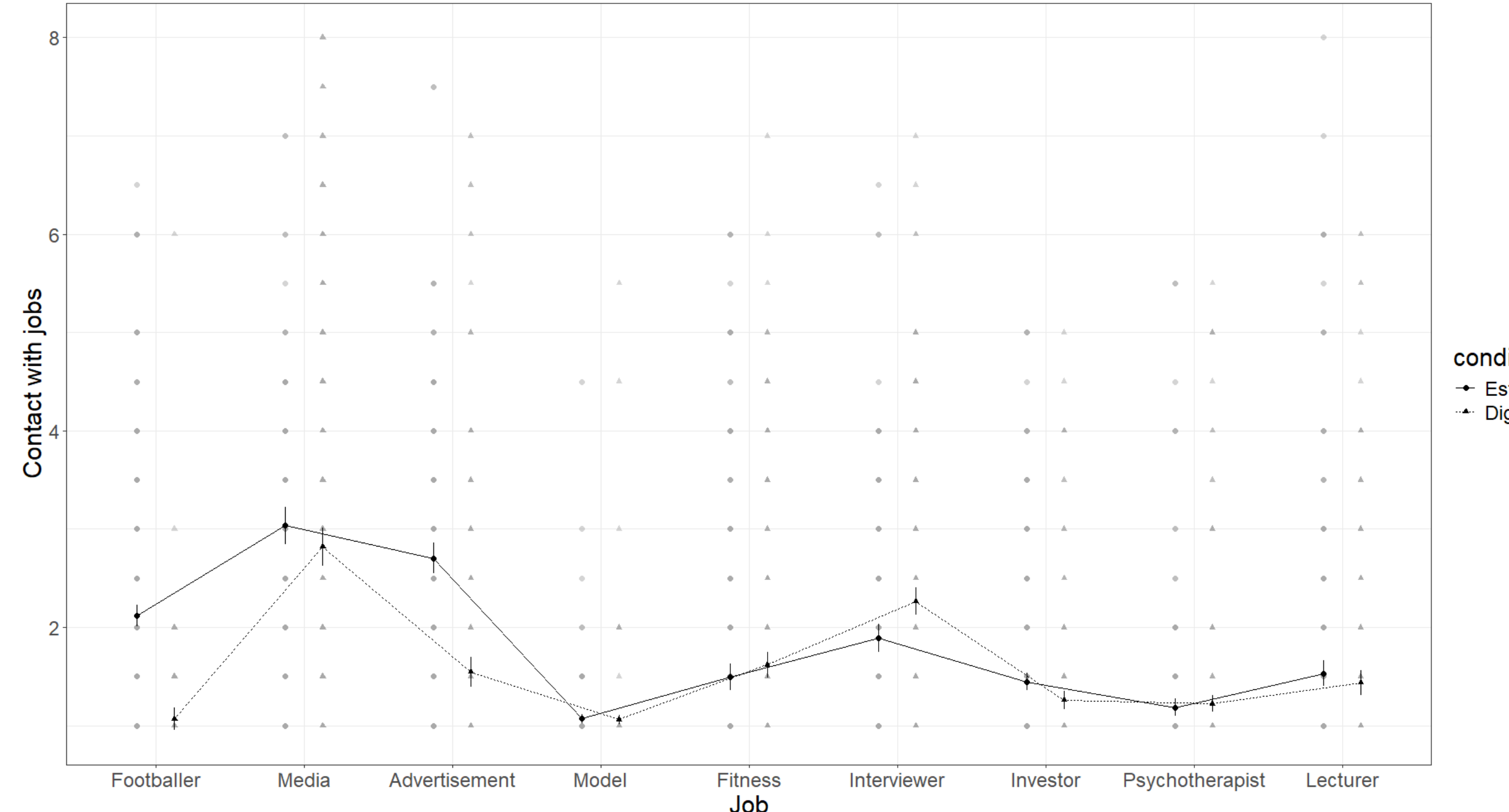

**Figure S11**

*Experiment 2: Association between openness with and attitudes towards jobs, separated by condition and job type*

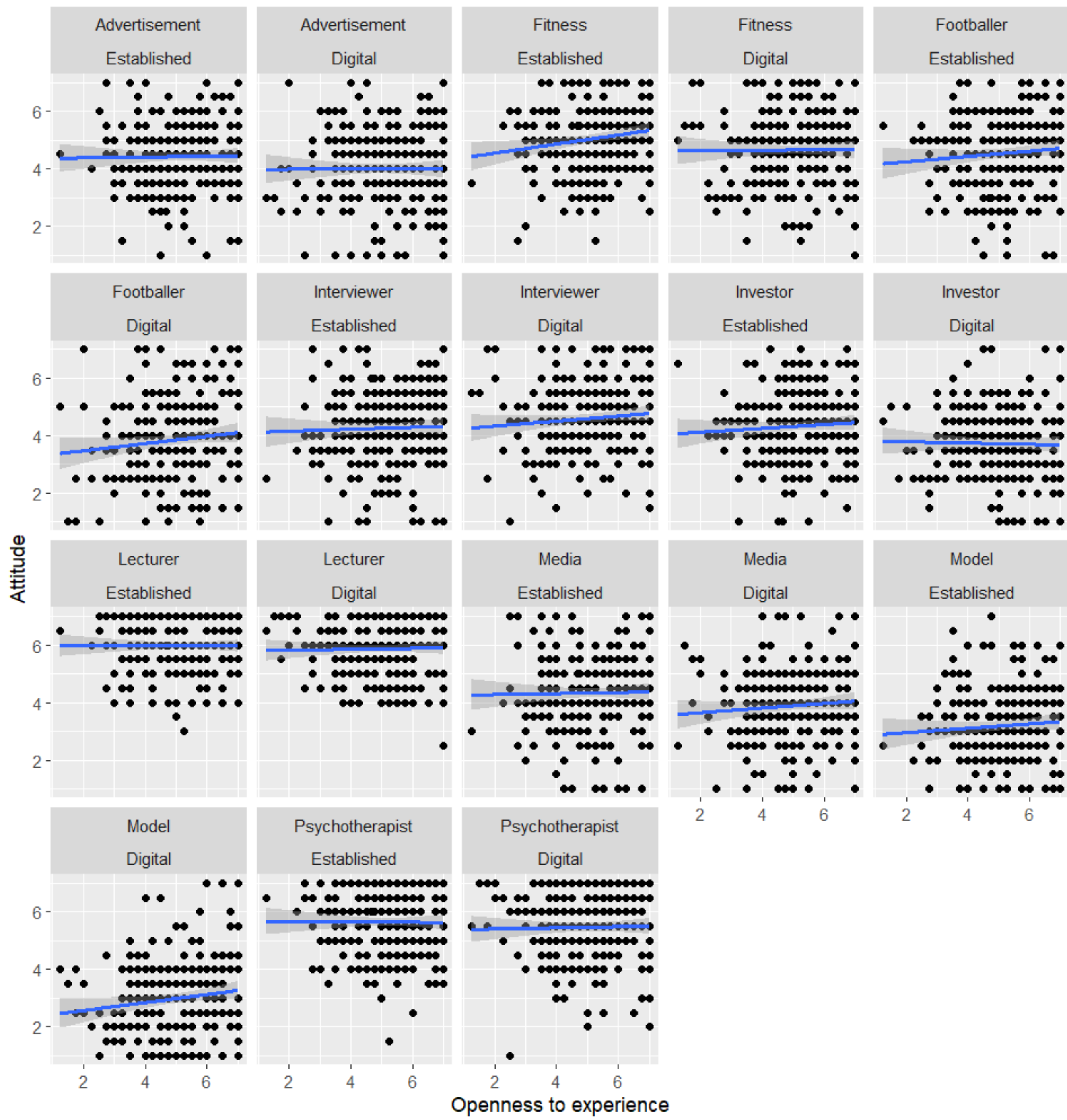

**Figure S12**

*Experiment 2: Association between attitudes towards digitalisation and attitudes towards jobs, separated by condition and job type*

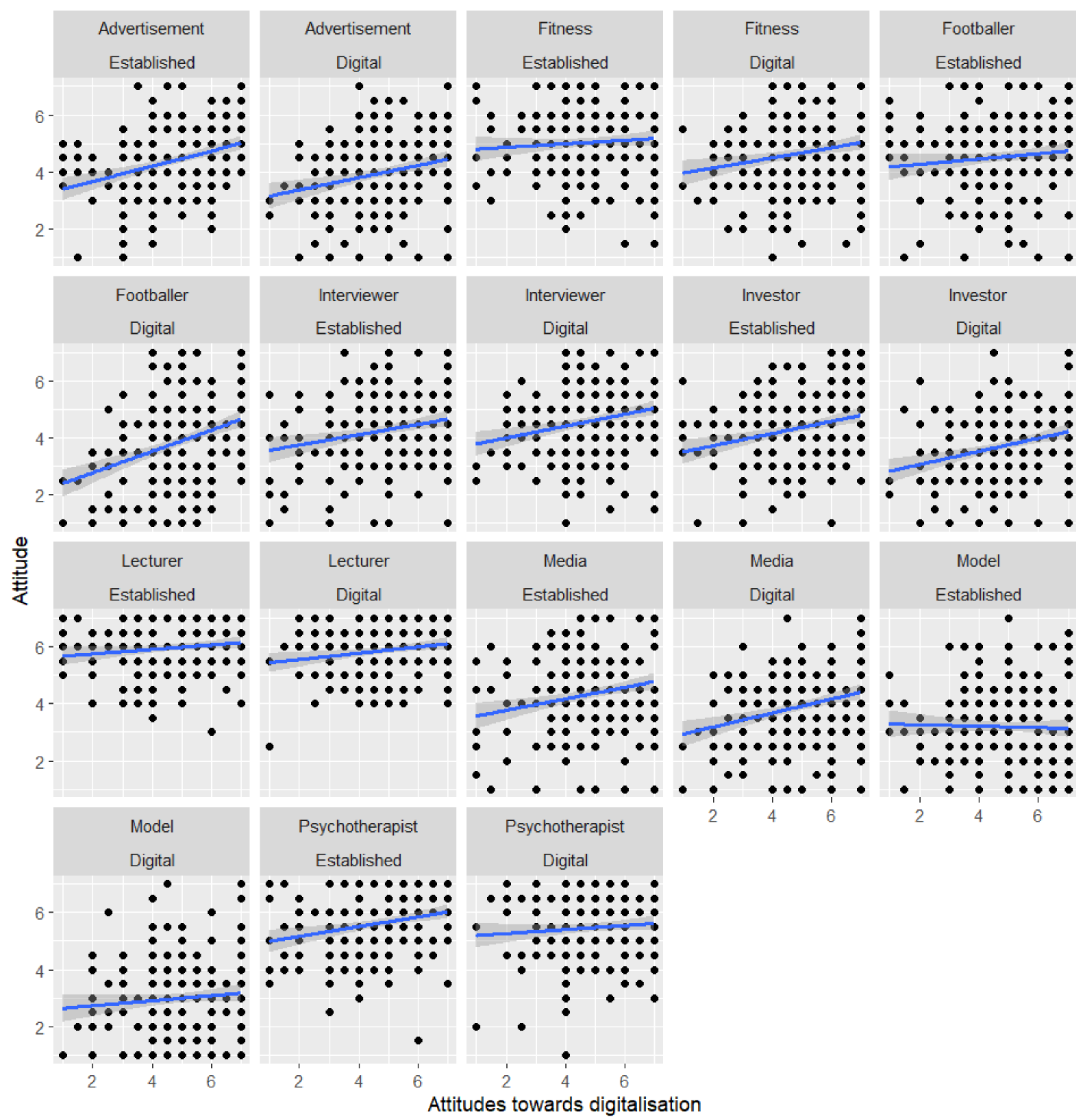

**Figure S13**

*Experiment 2: Association between political orientation and attitudes towards jobs, separated by condition and job type*

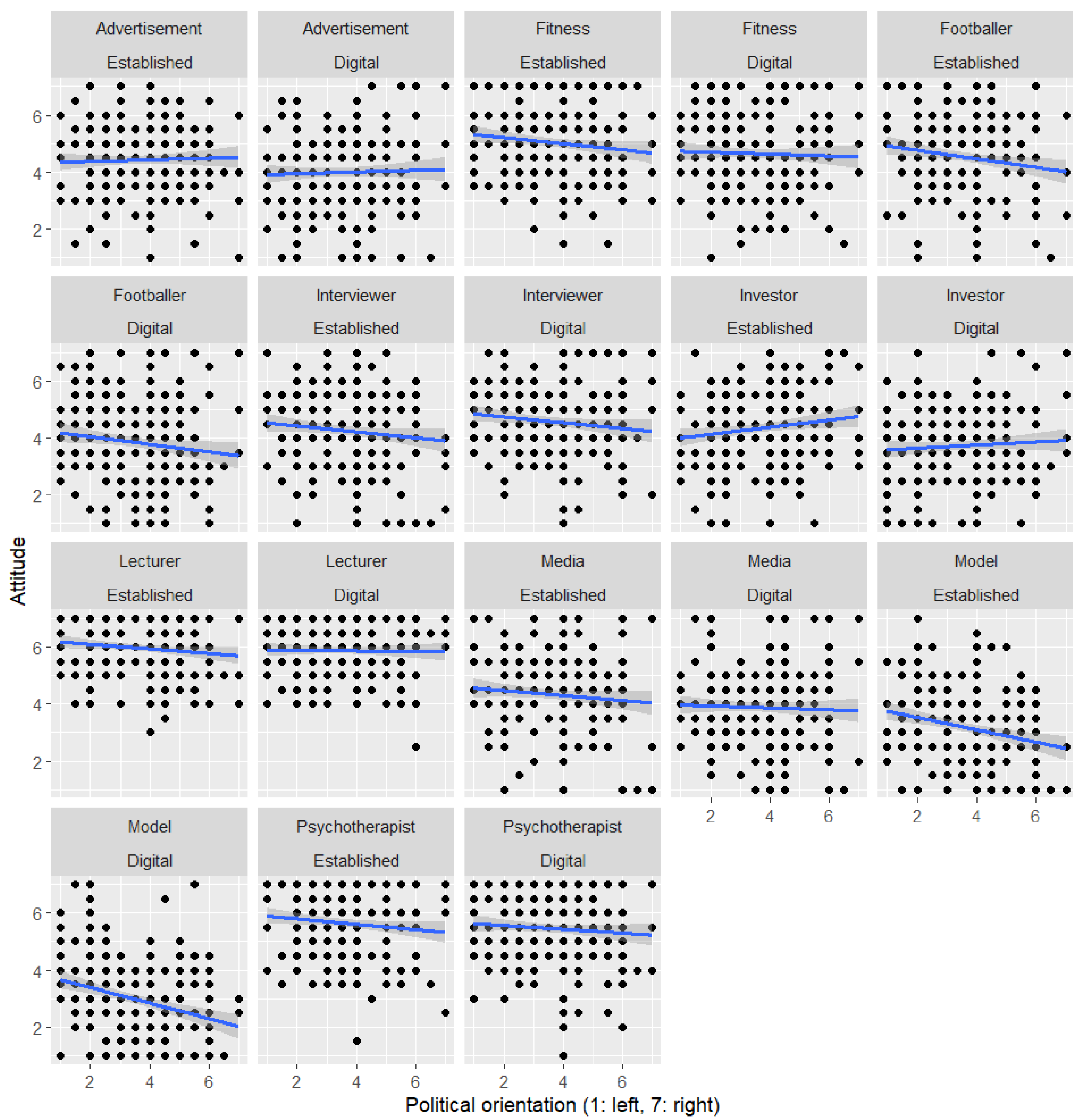

**Figure S14**

*Experiment 2: Association between age and attitudes towards jobs, separated by condition and job type*

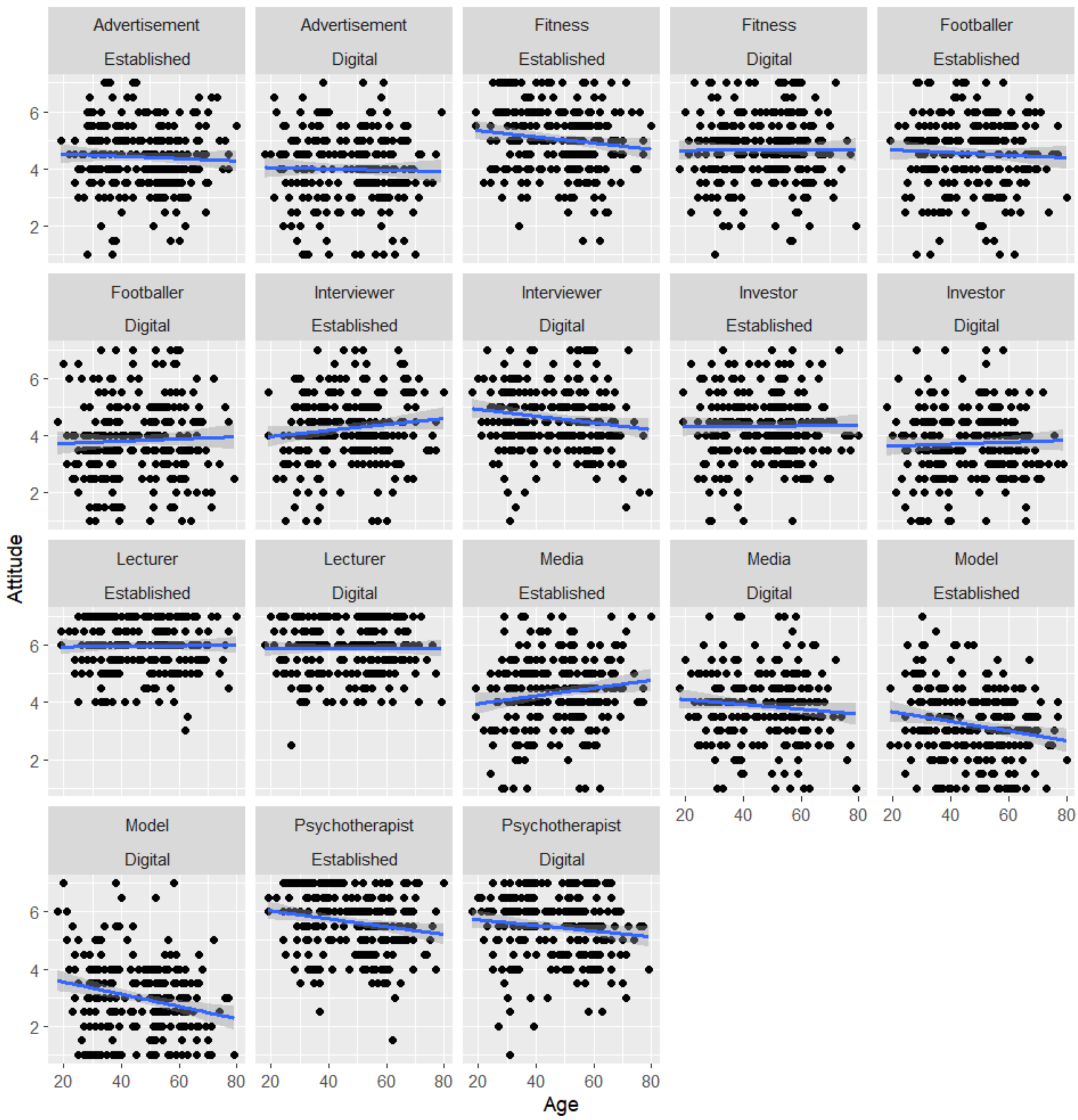

**Table S15**

*Consolidated Indirect Effects from Mediation Analyses: Standardized indirect effects (β) with bootstrapped 95% confidence intervals, by job type, mediator, and experiment*

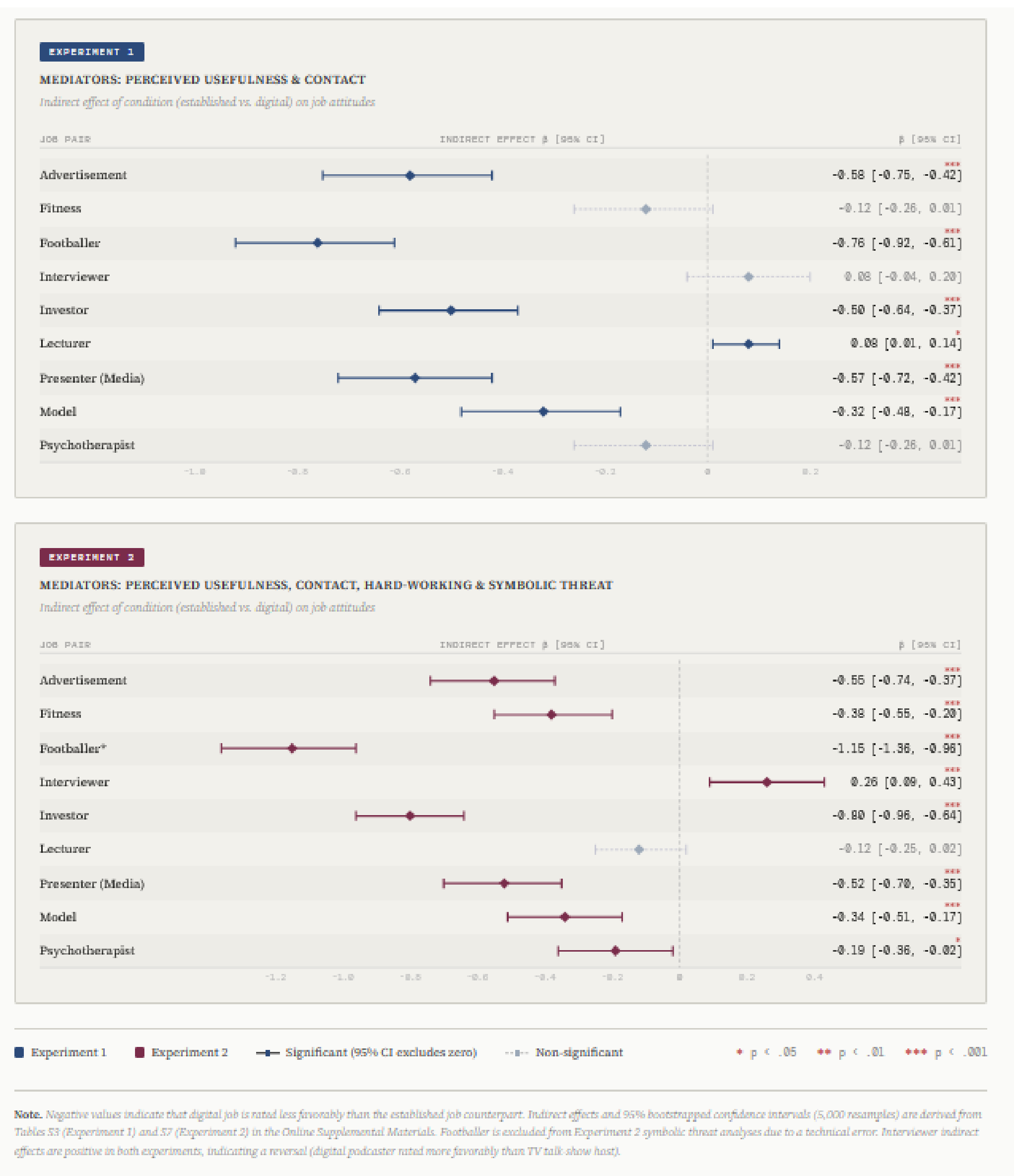

EXPERIMENT 1

MEDIATORS: PERCEIVED USEFULNESS & CONTACT

*Indirect effect of condition (established vs. digital) on job attitudes*

| JOB PAIR | β [95% CI] |
| --- | --- |
| Advertisement | -0.58 [-0.75, -0.42]*** |
| Fitness | -0.12 [-0.26, 0.01] |
| Footballer | -0.76 [-0.92, -0.61]*** |
| Interviewer | 0.08 [-0.04, 0.20] |
| Investor | -0.50 [-0.64, -0.37]*** |
| Lecturer | 0.08 [0.01, 0.14]* |
| Presenter (Media) | -0.57 [-0.72, -0.42]*** |
| Model | -0.32 [-0.48, -0.17]*** |
| Psychotherapist | -0.12 [-0.26, 0.01] |

EXPERIMENT 2

MEDIATORS: PERCEIVED USEFULNESS, CONTACT, HARD-WORKING & SYMBOLIC THREAT

*Indirect effect of condition (established vs. digital) on job attitudes*

| JOB PAIR | β [95% CI] |
| --- | --- |
| Advertisement | -0.55 [-0.74, -0.37]*** |
| Fitness | -0.38 [-0.55, -0.20]*** |
| Footballer* | -1.15 [-1.36, -0.96]*** |
| Interviewer | 0.26 [0.09, 0.43]*** |
| Investor | -0.80 [-0.96, -0.64]*** |
| Lecturer | -0.12 [-0.25, 0.02] |
| Presenter (Media) | -0.52 [-0.70, -0.35]*** |
| Model | -0.34 [-0.51, -0.17]*** |
| Psychotherapist | -0.19 [-0.36, -0.02]* |

Experiment 1 · Experiment 2 · Significant (95% CI excludes zero) · Non-significant · * $p < .05$ ** $p < .01$ *** $p < .001$

**Note.** *Negative values indicate that digital job is rated less favorably than the established job counterpart. Indirect effects and 95% bootstrapped confidence intervals (5,000 resamples) are derived from Tables S3 (Experiment 1) and S7 (Experiment 2) in the Online Supplemental Materials. Footballer is excluded from Experiment 2 symbolic threat analyses due to a technical error. Interviewer indirect effects are positive in both experiments, indicating a reversal (digital podcaster rated more favorably than TV talk show host).*

**Table S5**

*Experiment 2: Results of five 2 (condition) x 9 (jobs) mixed-ANOVAs: Main effects and interactions*

| | Condition | | Job | | Interaction | |
|---|---|---|---|---|---|---|
| | $F$ | $\eta_G^2$ | $F$ | $\eta_G^2$ | $F$ | $\eta_G^2$ |
| Attitudes | 17.77*** | .02 | 415.18*** | .31 | 13.35*** | .02 |
| Usefulness | 39.88*** | .03 | 861.98*** | .50 | 33.71*** | .04 |
| Hard working | 32.55*** | .03 | 186.21*** | .16 | 29.53*** | .03 |
| Symbolic^ | 33.03*** | .03 | 83.90*** | .07 | 5.36*** | .01 |
| Contact | 28.73*** | .01 | 187.44*** | .22 | 37.94*** | .06 |

*Note*. ^did not include footballer because of a technical error. *$p$ < .05, **$p$ < .01, ***$p$ < .001

**Table S6**

*Experiment 2: Mean comparisons of condition (established vs digital) separated by job type and dependent variable*

| | | Established | | Digital | | | | | |
|---|---|---|---|---|---|---|---|---|---|
| DV | Job | *M* | *SD* | *M* | *SD* | *t-value* | *p-value* | *d* | *U3* |
| Attitude | Advertisement | 4.42 | 1.13 | 3.99 | 1.25 | 4.05 | 0.0001 | 0.36 | 64.15 |
| Attitude | Fitness | 5.04 | 1.16 | 4.65 | 1.22 | 3.66 | 0.0003 | 0.33 | 62.81 |
| Attitude | Footballer | 4.53 | 1.26 | 3.85 | 1.39 | 5.75 | 0.0000 | 0.51 | 69.64 |
| Attitude | Interviewer | 4.26 | 1.23 | 4.59 | 1.17 | -3.09 | 0.0021 | -0.28 | 39.1 |
| Attitude | Investor | 4.33 | 1.19 | 3.73 | 1.17 | 5.70 | 0.0000 | 0.51 | 69.5 |
| Attitude | Lecturer | 5.97 | 0.87 | 5.87 | 0.92 | 1.25 | 0.2125 | 0.11 | 54.44 |
| Attitude | Media | 4.34 | 1.26 | 3.89 | 1.24 | 4.02 | 0.0001 | 0.36 | 64.03 |
| Attitude | Model | 3.20 | 1.28 | 2.99 | 1.31 | 1.80 | 0.0730 | 0.16 | 56.38 |
| Attitude | Psychotherapist | 5.65 | 1.09 | 5.47 | 1.14 | 1.80 | 0.0725 | 0.16 | 56.39 |
| Contact | Advertisement | 2.70 | 1.32 | 1.55 | 1.16 | 10.42 | 0.0000 | 0.93 | 82.43 |
| Contact | Fitness | 1.49 | 1.07 | 1.62 | 1.09 | -1.29 | 0.1991 | -0.11 | 45.42 |
| Contact | Footballer | 2.12 | 1.21 | 1.07 | 0.38 | 13.14 | 0.0000 | 1.18 | 88 |
| Contact | Interviewer | 1.89 | 0.93 | 2.26 | 1.30 | -3.69 | 0.0002 | -0.33 | 37.05 |
| Contact | Investor | 1.45 | 0.82 | 1.26 | 0.64 | 2.86 | 0.0045 | 0.26 | 60.09 |
| Contact | Lecturer | 1.53 | 1.12 | 1.44 | 0.93 | 1.02 | 0.3079 | 0.09 | 53.64 |
| Contact | Media | 3.03 | 1.18 | 2.82 | 1.85 | 1.54 | 0.1243 | 0.14 | 55.48 |
| Contact | Model | 1.07 | 0.31 | 1.06 | 0.40 | 0.25 | 0.8036 | 0.02 | 50.89 |
| Contact | Psychotherapist | 1.19 | 0.63 | 1.22 | 0.76 | -0.61 | 0.5438 | -0.05 | 47.83 |
| Hard-work | Advertisement | 4.11 | 0.76 | 3.72 | 0.89 | 5.18 | 0.0000 | 0.46 | 67.86 |
| Hard-work | Fitness | 4.25 | 0.70 | 4.04 | 0.74 | 3.19 | 0.0015 | 0.29 | 61.23 |
| Hard-work | Footballer | 4.09 | 0.79 | 3.30 | 1.02 | 9.69 | 0.0000 | 0.87 | 80.68 |
| Hard-work | Interviewer | 3.92 | 0.84 | 3.98 | 0.83 | -0.83 | 0.4048 | -0.07 | 47.03 |
| Hard-work | Investor | 3.96 | 0.84 | 3.31 | 0.94 | 8.11 | 0.0000 | 0.73 | 76.58 |
| Hard-work | Lecturer | 4.42 | 0.61 | 4.48 | 0.58 | -1.20 | 0.2295 | -0.11 | 45.72 |
| Hard-work | Media | 3.99 | 0.82 | 3.57 | 0.93 | 5.32 | 0.0000 | 0.48 | 68.28 |
| Hard-work | Model | 3.37 | 1.05 | 3.01 | 1.04 | 3.87 | 0.0001 | 0.35 | 63.55 |
| Hard-work | Psychotherapist | 4.31 | 0.71 | 4.29 | 0.68 | 0.29 | 0.7713 | 0.03 | 51.04 |
| Symbolic | Advertisement | 1.93 | 0.95 | 2.47 | 1.12 | -5.75 | 0.0000 | -0.51 | 30.35 |
| Symbolic | Fitness | 1.59 | 0.81 | 2.02 | 0.98 | -5.32 | 0.0000 | -0.48 | 31.72 |
| Symbolic | Footballer | | | | | | | | |
| Symbolic | Interviewer | 1.95 | 0.98 | 2.00 | 0.95 | -0.67 | 0.5004 | -0.06 | 47.6 |
| Symbolic | Investor | 2.07 | 1.03 | 2.41 | 1.04 | -3.74 | 0.0002 | -0.33 | 36.91 |
| Symbolic | Lecturer | 1.35 | 0.61 | 1.83 | 0.91 | -6.89 | 0.0000 | -0.62 | 26.88 |
| Symbolic | Media | 2.00 | 1.05 | 2.37 | 1.05 | -3.89 | 0.0001 | -0.35 | 36.39 |
| Symbolic | Model | 2.19 | 1.06 | 2.49 | 1.14 | -2.99 | 0.0030 | -0.27 | 39.46 |
| Symbolic | Psychotherapist | 1.44 | 0.68 | 1.85 | 0.96 | -5.61 | 0.0000 | -0.50 | 30.8 |
| Usefulness | Advertisement | 3.76 | 1.35 | 3.18 | 1.43 | 4.66 | 0.0000 | 0.42 | 66.15 |
| Usefulness | Fitness | 4.90 | 1.26 | 4.48 | 1.26 | 3.70 | 0.0002 | 0.33 | 62.98 |
| Usefulness | Footballer | 3.38 | 1.37 | 2.18 | 1.19 | 10.41 | 0.0000 | 0.93 | 82.4 |
| Usefulness | Interviewer | 3.48 | 1.40 | 3.95 | 1.32 | -3.87 | 0.0001 | -0.35 | 36.46 |
| Usefulness | Investor | 3.60 | 1.50 | 2.38 | 1.29 | 9.78 | 0.0000 | 0.87 | 80.91 |
| Usefulness | Lecturer | 6.09 | 0.93 | 5.93 | 0.89 | 1.96 | 0.0503 | 0.18 | 56.97 |
| Usefulness | Media | 3.75 | 1.39 | 3.14 | 1.41 | 4.84 | 0.0000 | 0.43 | 66.74 |
| Usefulness | Model | 2.17 | 1.16 | 1.94 | 1.06 | 2.32 | 0.0208 | 0.21 | 58.22 |
| Usefulness | Psychotherapist | 5.81 | 1.09 | 5.58 | 1.18 | 2.21 | 0.0279 | 0.20 | 57.82 |

*Note*. *Note*. d: Cohen's d (standardized mean difference). U3: Cohen's U3 (percentage of participants who hold more favourable attitudes towards the established job compared to the average of the digital job). We report also Cohen's U3 because it is perceived as more informative than several other effect sizes (Hanel & Mehler, 2019). All p-values are uncorrected but two-tailed. Attitude: Attitudes towards job, hard-work: How hard working are people doing the job perceived, symbolic: symbolic threat. All p-values are uncorrected.

**Table S7**

*Experiment 2: Mediation analyses of condition on attitudes with usefulness and contact as mediator (bootstrapped 95%-CIs)*

| | c: total effect | c': Direct effect | Indirect effect [95%-CI] | $R^2$ |
|---|---|---|---|---|
| Advertisement | -.43*** | .11 | -.55 [-.74, -.37] | .65*** |
| Fitness | -.39*** | -.01 | -.38 [-.55, -.20] | .69*** |
| Footballer^ | -.68*** | .47*** | -1.15 [-1.36, -.96] | .57*** |
| Interviewer | .33** | .07 | .26 [.09, .43] | .64*** |
| Investor | -.60*** | .20* | -.80 [-.96, -.64] | .54*** |
| Lecturer | -.10 | .02 | -.12 [-.25, .02] | .63*** |
| Presenter (media) | -.45*** | .08 | -.52 [-.70, -.35] | .67*** |
| Model | -.21 | .13 | -.34 [-.51, -.17] | .56*** |
| Psychotherapist | -.18 | .01 | -.19 [-.36, -.02] | .69*** |

*Note.* Bootstrapped 95%-CIs are based on 5000 resamples. If total effect is negative, attitudes towards the established version of the job are more positive (cf. Table 6 for mean differences). ^did not include symbolic threat because of a technical error. $^{*}p < .05$, $^{**}p < .01$, $^{***}p < .001$

**Table S8**

*Experiment 2: Correlations between attitudes and the four moderators by job and condition*

| | | Openness | | Digital | | Political | | Age | |
|---|---|---|---|---|---|---|---|---|---|
| job | condition | *r* | *p* | *r* | *p* | *r* | *p* | *r* | *p* |
| Advertisement | Established | .01 | .8230 | .33 | .0000 | .04 | .5820 | -.05 | .4470 |
| Advertisement | Digital | .00 | .9790 | .24 | .0001 | .03 | .6130 | -.03 | .6950 |
| Fitness | Established | .16 | .0092 | .08 | .2390 | -.13 | .0435 | -.13 | .0460 |
| Fitness | Digital | .01 | .9290 | .20 | .0013 | -.04 | .5410 | .00 | .9400 |
| Footballer | Established | .09 | .1690 | .10 | .1060 | -.17 | .0088 | -.05 | .4400 |
| Footballer | Digital | .11 | .0750 | .37 | .0000 | -.13 | .0331 | .04 | .5440 |
| Interviewer | Established | .03 | .6010 | .21 | .0011 | -.12 | .0653 | .11 | .0707 |
| Interviewer | Digital | .09 | .1420 | .25 | .0001 | -.12 | .0504 | -.14 | .0268 |
| Investor | Established | .07 | .2900 | .25 | .0001 | .15 | .0202 | .01 | .9030 |
| Investor | Digital | -.02 | .7120 | .28 | .0000 | .07 | .2970 | .04 | .5370 |
| Lecturer | Established | .00 | .9600 | .13 | .0483 | -.13 | .0444 | .02 | .8070 |
| Lecturer | Digital | .02 | .7430 | .17 | .0076 | -.02 | .7660 | .00 | .9480 |
| Media | Established | .02 | .7860 | .22 | .0004 | -.10 | .1330 | .15 | .0215 |
| Media | Digital | .08 | .2030 | .28 | .0000 | -.04 | .5470 | -.10 | .1360 |
| Model | Established | .07 | .2610 | -.03 | .6690 | -.24 | .0002 | -.18 | .0051 |
| Model | Digital | .13 | .0370 | .09 | .1380 | -.30 | .0000 | -.23 | .0003 |
| Psychotherapist | Established | -.01 | .9120 | .22 | .0005 | -.12 | .0571 | -.17 | .0073 |
| Psychotherapist | Digital | .02 | .7390 | .09 | .1770 | -.08 | .1920 | -.12 | .0603 |

*Note*. Correlations between attitudes towards the jobs and each of the four moderators. For instance, the correlation between attitudes towards established advertisement job and openness to new experience is $r = .10$, $p = .1050$ (top left of table above). Openness: openness to new experience, digital: attitudes towards digitalization, political: political orientation, age: Age of participants.